\documentclass[preprint,twocolumn]{aastex63}
\usepackage{float, bm, graphicx, amsmath, morefloats}
\usepackage[caption=false]{subfig}
\PassOptionsToPackage{hyphens}{url}\usepackage{hyperref}
\usepackage{xurl}

\submitjournal{ApJ}

\shorttitle{Afterglows}
\shortauthors{Ho et al.}

\turnoffeditone

\newcommand{\swift}{{\em Swift}}

\newcommand{\erg}{\mbox{$\rm erg$}}

\newcommand{\km}{\mbox{$\rm km$}}

\newcommand{\pmpc}{\mbox{$\rm Mpc^{-1}$}}

\newcommand{\pcmsq}{\mbox{$\rm cm^{-2}$}}

\newcommand{\degsq}{\mbox{$\rm deg^{2}$}}

\newcommand{\pcmcub}{\mbox{$\rm cm^{-3}$}}

\newcommand{\pgpccub}{\mbox{$\rm Gpc^{-3}$}}

\newcommand{\msol}{\mbox{$\rm M_\odot$}}

\newcommand{\days}{\mbox{$\rm d$}}

\newcommand{\psec}{\mbox{$\rm s^{-1}$}}
\newcommand{\pyr}{\mbox{$\rm yr^{-1}$}}
\newcommand{\pday}{\mbox{$\rm d^{-1}$}}

\newcommand{\phz}{\mbox{$\rm Hz^{-1}$}}

\shorttitle{ZTF Afterglows}

\begin{document}

\title{Cosmological Fast Optical Transients with the Zwicky Transient Facility:\\
A Search for Dirty Fireballs}

\correspondingauthor{Anna Y. Q. Ho}
\email{annayqho@cornell.edu}

\author[0000-0002-9017-3567]{Anna Y. Q.~Ho}
\affiliation{Miller Institute for Basic Research in Science, 468 Donner Lab, Berkeley, CA 94720, USA}
\affiliation{Department of Astronomy, University of California, Berkeley, CA, 94720-3411, USA}
\affiliation{Lawrence Berkeley National Laboratory, 1 Cyclotron Road, MS 50B-4206, Berkeley, CA 94720, USA}
\affiliation{Department of Astronomy, Cornell University, Ithaca, NY 14853, USA}

\author[0000-0001-8472-1996]{Daniel A.~Perley}
\affiliation{Astrophysics Research Institute, Liverpool John Moores University, IC2, Liverpool Science Park, 146 Brownlow Hill, Liverpool L3 5RF, UK}

\author[0000-0001-6747-8509]{Yuhan Yao}
\affiliation{Cahill Center for Astrophysics, 
California Institute of Technology, MC 249-17, 
1200 E California Boulevard, Pasadena, CA, 91125, USA}

\author[0000-0002-2208-2196]{Dmitry Svinkin}
\affiliation{Ioffe Institute, 26 Politekhnicheskaya, St Petersburg 194021, Russia}

\author[0000-0001-7717-5085]{A.~de Ugarte Postigo}
\affiliation{Artemis, Observatoire de la Côte d'Azur, Universit\'{e} Côte d'Azur, CNRS, 06304 Nice, France
}

\author[0000-0001-7097-8360]{R. A. Perley}
\affiliation{National Radio Astronomy Observatory, P.O. Box O, Socorro, NM 87801, USA}

\author[0000-0003-2902-3583]{D.~Alexander~Kann}
\affiliation{Instituto de Astrof\'{i}sica de Andaluc\'{i}a (IAA-CSIC), Glorieta de la
Astronom\'{i}a s/n, 18008 Granada, Spain}

\author[0000-0002-2942-3379]{Eric Burns}
\affiliation{Department of Physics \& Astronomy, Louisiana State University, Baton Rouge, LA 70803, USA}

% ALPHABETIZED

\author[0000-0003-3768-7515]{Igor Andreoni}
\affil{Joint Space-Science Institute, University of Maryland, College Park, MD 20742, USA.}
\affil{Department of Astronomy, University of Maryland, College Park, MD 20742, USA.}
\affil{Astrophysics Science Division, NASA Goddard Space Flight Center, Mail Code 661, Greenbelt, MD 20771, USA}

\author[0000-0001-8018-5348]{Eric C. Bellm}
\affiliation{DIRAC Institute, Department of Astronomy, University of Washington, 3910 15th Avenue NE, Seattle, WA 98195, USA}

\author[0000-0001-9935-8106]{Elisabetta~Bissaldi}
\affiliation{Dipartimento di Fisica ``M. Merlin" dell'Universit\`a e del Politecnico di Bari, I-70126 Bari, Italy}
\affiliation{Istituto Nazionale di Fisica Nucleare, Sezione di Bari, I-70126 Bari, Italy}

 \author[0000-0002-7777-216X]{Joshua S. Bloom}
\affiliation{Department of Astronomy, University of California, Berkeley, CA, 94720-3411, USA}
\affiliation{Lawrence Berkeley National Laboratory, 1 Cyclotron Road, MS 50B-4206, Berkeley, CA 94720, USA}

\author[0000-0001-5955-2502]{Thomas G. Brink}
\affiliation{Department of Astronomy, University of California, Berkeley, CA, 94720-3411, USA}

\author[0000-0002-5884-7867]{Richard Dekany}
\affiliation{Caltech Optical Observatories, California Institute of Technology, Pasadena, CA  91125}

\author{Andrew J. Drake}
\affiliation{Cahill Center for Astrophysics, 
California Institute of Technology, MC 249-17, 
1200 E California Boulevard, Pasadena, CA, 91125, USA}

\author[0000-0001-6991-7616]{Jos\'{e} Feliciano Ag\"{u}\'{i} Fern\'{a}ndez}
\affiliation{Instituto de Astrof\'{i}sica de Andaluc\'{i}a (IAA-CSIC), Glorieta de la
Astronom\'{i}a s/n, 18008 Granada, Spain}

\author[0000-0003-3460-0103]{Alexei V. Filippenko}
\affiliation{Department of Astronomy,
University of California,
Berkeley, CA, 94720-3411, USA}

\author[0000-0002-1153-6340]{Dmitry Frederiks}
\affiliation{Ioffe Institute, 26 Politekhnicheskaya, St Petersburg 194021, Russia}

\author[0000-0002-3168-0139]{Matthew~J.~Graham}
\affiliation{Cahill Center for Astrophysics, 
California Institute of Technology, MC 249-17, 
1200 E California Boulevard, Pasadena, CA, 91125, USA}

\author[0000-0001-9556-7576]{Boyan~A.~Hristov}
\affiliation{Center for Space Plasma and Aeronomic Research (CSPAR), University of Alabama in Huntsville, Huntsville, AL 35899, USA}

\author[0000-0002-5619-4938]{Mansi~M.~Kasliwal}
\affiliation{Cahill Center for Astrophysics, 
California Institute of Technology, MC 249-17, 
1200 E California Boulevard, Pasadena, CA, 91125, USA}

\author[0000-0001-5390-8563]{S.~R.~Kulkarni}
\affiliation{Cahill Center for Astrophysics, 
California Institute of Technology, MC 249-17, 
1200 E California Boulevard, Pasadena, CA, 91125, USA}

\author[0000-0003-0871-4641]{Harsh Kumar}
\affiliation{Department of Physics, Indian Institute of Technology Bombay, Powai, Mumbai-400076, India}
\affiliation{LSSTC Data Science Fellow 2018}

\author[0000-0003-2451-5482]{Russ R. Laher}
\affiliation{IPAC, California Institute of Technology, 1200 E. California Blvd, Pasadena, CA 91125, USA}

\author[0000-0002-3942-8341]{Alexandra L. Lysenko}
\affiliation{Ioffe Institute, 26 Politekhnicheskaya, St Petersburg 194021, Russia}

\author{Bagrat~Mailyan}
\affiliation{Center for Astro, Particle, and Planetary Physics, New York University Abu Dhabi, Abu Dhabi, UAE}

\author[0000-0002-0380-0041]{Christian~Malacaria}
\affiliation{Universities Space Research Association, NSSTC, 320 Sparkman Drive, Huntsville, AL 35805, USA}
            
\author[0000-0001-9515-478X]{A.~A.~Miller}
\affiliation{Center for Interdisciplinary Exploration and Research in
             Astrophysics (CIERA) and Department of Physics and Astronomy,
             Northwestern University,
             1800 Sherman Road, Evanston, IL 60201, USA}

\author[0000-0002-6269-0452]{S.~Poolakkil}
\affiliation{Department of Space Science, University of Alabama in Huntsville, Huntsville, AL 35899, USA}
\affiliation{Center for Space Plasma and Aeronomic Research, University of Alabama in Huntsville, Huntsville, AL 35899, USA}
    
\author[0000-0002-0387-370X]{Reed Riddle}
\affiliation{Caltech Optical Observatories, California Institute of Technology, Pasadena, CA  91125}

\author[0000-0001-9477-5437]{Anna Ridnaia}
\affiliation{Ioffe Institute, 26 Politekhnicheskaya, St Petersburg 194021, Russia}

\author[0000-0001-7648-4142]{Ben Rusholme}
\affiliation{IPAC, California Institute of Technology, 1200 E. California
             Blvd, Pasadena, CA 91125, USA}

\author{Volodymyr Savchenko}
\affiliation{ISDC, Department of Astronomy, University of Geneva, Chemin d'Ecogia, 16 CH-1290 Versoix, Switzerland}

\author[0000-0003-1546-6615]{Jesper Sollerman}
\affiliation{The Oskar Klein Centre, Department of Astronomy, Stockholm University, AlbaNova, SE-10691 Stockholm, Sweden}

\author[0000-0002-7978-7648]{Christina Th\"{o}ne}
\affiliation{Astronomical Institute of the Czech Academy of Sciences (ASU-CAS), Fri\v
cova 298, 251 65 Ond\v rejov, CZ}

\author[0000-0003-0292-6221]{Anastasia Tsvetkova}
\affiliation{Ioffe Institute, 26 Politekhnicheskaya, St Petersburg 194021, Russia}

\author[0000-0002-0076-5228]{Mikhail Ulanov}
\affiliation{Ioffe Institute, 26 Politekhnicheskaya, St Petersburg 194021, Russia}

\author[0000-0002-0221-5916]{Andreas~von~Kienlin}
\affiliation{Max-Planck Institut f\"ur extraterrestrische Physik, D-85748 Garching, Germany}

\begin{abstract}

Dirty fireballs are a hypothesized class of relativistic massive-star explosions with an initial Lorentz factor $\Gamma_\mathrm{init}$ below the $\Gamma_\mathrm{init}\sim 100$ required to produce a long-duration gamma-ray burst (LGRB),
but which could still produce optical emission resembling LGRB afterglows.
Here we present the results of a search for on-axis optical afterglows using the Zwicky Transient Facility (ZTF).
Our search yielded seven optical transients that resemble on-axis LGRB afterglows in terms of their red colors ($g-r>0\,$mag), faint host galaxy ($r>23\,$mag), and rapid fading ($dr/dt>1\,$mag\,\pday).
Spectroscopy of the transient emission within a few days of discovery established cosmological distances (redshift $z=0.876$ to 2.9) for six events, tripling the number of afterglows with redshift measurements discovered by optical surveys without a $\gamma$-ray trigger.
A likely associated LGRB (GRB\,200524A, GRB\,210204A, GRB\,210212B, GRB\,210610B) was identified for four events (ZTF\,20abbiixp/AT\,2020kym, ZTF\,21aagwbjr/AT\,2021buv, ZTF\,21aakruew/AT\,2021cwd, ZTF\,21abfmpwn/AT\,2021qbd) post facto,
while three (ZTF\,20aajnksq/AT\,2020blt, ZTF\,21aaeyldq/AT\,2021any,  ZTF\,21aayokph/AT\,2021lfa) had no detected LGRB counterpart.
The simplest explanation for the three ``orphan'' events is that they were regular LGRBs missed by high-energy satellites owing to detector sensitivity and duty cycle, although it is possible that they were intrinsically subluminous in $\gamma$-rays or viewed slightly off-axis.
We rule out a scenario in which dirty fireballs have a similar energy per solid angle to LGRBs and are an order of magnitude more common.
In addition, we set the first direct constraint on the ratio of the opening angles of the material producing $\gamma$-rays and the material producing early optical afterglow emission, finding that they must be comparable.

\end{abstract}

\section{Introduction}
\label{sec:intro}

Decades of observations of long-duration $\gamma$-ray bursts (LGRBs) and their associated afterglows have revealed that in the deaths of some massive ($M > 10\,M_\odot$) stripped-envelope stars, the newborn compact object can couple $10^{51}\,$erg of energy to ultrarelativistic (initial Lorentz factor $\Gamma_\mathrm{init} \gg 100$) material,
and that this phenomenon preferentially occurs in low-metallicity environments \citep{Kouveliotou2012}.
In the traditional LGRB model \citep{Piran2004,Meszaros2006,Kumar2015},
the outflow is collimated into a jet a few degrees wide as it tunnels through the star.
Viewed on-axis, the observer sees a seconds-long burst of $\gamma$-rays from collisions within the jet, then emission from X-ray to radio wavelengths called the ``afterglow'' \citep{Meszaros1997,Sari1998,vanParadijs2000,Panaitescu2002,Greiner2011} on timescales of minutes to years.
Thousands of afterglows have been detected to date\footnote{\url{https://www.mpe.mpg.de/~jcg/grbgen.html}},
almost all in targeted follow-up observations of discoveries by GRB satellites such as the \emph{High Energy Transient Explorer 2} (\emph{HETE-2}; \citealt{Ricker2003}), \emph{Fermi} \citep{Meegan2009}, and \emph{Swift} \citep{Gehrels2004}.

Despite the significant progress made by LGRB discoveries,
many important questions cannot be answered by relying on existing $\gamma$-ray satellites to discover relativistic outflows from collapsing stars.
In particular, the rate and angular structure of LGRB jets are unknown in large part because observed LGRBs are viewed almost entirely on-axis (or close to on-axis; \citealt{Ryan2015}), with only a few suggested exceptions \citep{Huang2004,RamirezRuiz2005,Butler2005,Kruhler2009}.
Relatedly, it is unknown whether ultrarelativistic speeds, which can only be attained with very small amounts of matter entrained in the jet (``low mass-loading''), are central to the phenomenon.
It has been hypothesized that mass-loaded jets (``dirty fireballs;'' \citealt{Paczynski1998,Dermer1999_dirty_fireballs}) are more common \citep{Huang2002}, but have gone unnoticed because their emission peaks at energies below the range of $\gamma$-ray detectors \citep{Dermer1999_dirty_fireballs}, instead appearing as X-ray flashes \citep{Heise2001,Zhang2004_simulation,Sakamoto2005,Soderberg2007}.

A promising approach to answering the above questions is to find relativistic outflows via their afterglow emission, without relying on a trigger from a $\gamma$-ray satellite.
Assuming that dirty fireballs result in successful relativistic outflows with a similar energy per solid angle to that of LGRBs, they should produce luminous and rapidly fading optical afterglows \citep{Huang2002,Rhoads2003}.
The afterglows from clean or dirty fireballs viewed off-axis ($\theta_\mathrm{obs} > 1/\Gamma_\mathrm{init}$) will eventually become visible to the observer as the jet decelerates and expands sideways, often referred to as ``orphan'' afterglows \citep{Rhoads1997,Perna1998,Dalal2002,Granot2002,Nakar2002,Totani2002}.
For a structured jet (see \citealt{Granot2010} for an overview),
orphan afterglows could also be detected from jets viewed within the initial opening angle of relativistic material, but outside the narrow high-Lorentz factor region emitting $\gamma$-rays; such events have been referred to as ``on-axis orphans'' \citep{Nakar2003}.

Finding optical afterglows without a GRB trigger is a longstanding goal of transient surveys, and prior to the Zwicky Transient Facility (ZTF; \citealt{Bellm2019_ztf,Graham2019}) had only been achieved three times.
Searches at X-ray \citep{Grindlay1999,Greiner2000,Khabibullin2012}, optical \citep{Rau2006,Berger2013,Ho2018,Huang2020}, and radio \citep{Levinson2002,GalYam2006} wavelengths have to contend with a large false-positive rate, particularly from stellar flares at optical and X-ray wavelengths, and from active galactic nuclei (AGNs) at radio wavelengths.
Nonetheless, two confirmed optical afterglows were serendipitously discovered via fading broadband afterglow emission:
iPTF14yb \citep{Cenko2015} and ATLAS17aeu \citep{Stalder2017,Bhalerao2017,Melandri2019} were discovered by the Palomar Transient Factory \citep{Law2009} and the ATLAS survey \citep{Tonry2018}, respectively, with LGRB counterparts discovered post facto.
A third optical event, PTF11agg \citep{Cenko2013},
resembled a GRB afterglow (rapidly fading optical emission, a long-lived scintillating radio counterpart, coincidence with a dwarf galaxy) yet had no identified high-energy counterpart, leading to suggestions that it might be a dirty fireball.
\edit1{At radio wavelengths, a promising candidate off-axis afterglow has been identified in VLA Sky Survey data \citep{Law2018,Mooley2022}.}

In the last few years, with the enhanced survey speed \citep{Bellm2016_grasp,Ofek2020} of ZTF, the discovery of optical afterglows without a GRB trigger has become routine.
As we discuss in this paper, ten afterglows have been discovered to date using ZTF data, five from the year 2021 alone.
Most were identified by human scanners within 12\,hr of the first ZTF detection, through dedicated searches for fast optical transients that make use of the transient's rise rate, fade rate, color, and contextual information such as the host galaxy \citep{Ho2020d,Andreoni2020,Andreoni2021}.
\edit1{Searches for very fast ($\ll$30\,s) optical transients accompanying fast radio bursts have also been conducted using ZTF data \citep{Andreoni2020_FRB}}.

\edit1{In this paper we present discovery and follow-up details for six of the ten ZTF afterglows (AT\,2020kym, AT\,2021any, AT\,2021buv, AT\,2021cwd, AT\,2021lfa, AT\,2021qbd), including two (AT\,2021any and AT\,2021lfa) with no detected GRB counterpart. 
We also present deep imaging with Keck at the position of the ZTF afterglow AT\,2020blt, the discovery of which was published in \citet{Ho2020b}.
We do not present new data on the remaining three afterglows, two of which (AT\,2020sev and AT\,2020yxz) had detected GRB counterparts and were published in \citet{Andreoni2021}, and one of which (AT\,2019pim) had no GRB counterpart and will be published separately (Perley et al in prep.).}
We describe our search strategy and give an overview of the sample in \S\ref{sec:observations}.
In \S\ref{sec:comparison} we compare the multiwavelength properties of the ZTF afterglows to the cosmological LGRB population.
\S\ref{sec:discussion} discusses the implications for dirty fireballs, jet collimation, and the prevalence of relativistic jets in collapsing massive stars.
We summarize in \S\ref{sec:conclusions} and discuss how to make future progress.

Throughout the paper we use the term ``afterglow'' to refer to cosmological fast optical transients whose observed properties strongly resemble those of GRB afterglows. In addition, for brevity we use the term ``orphan afterglow'' to refer to afterglows with no associated detected GRB.
However, as discussed in \S\ref{sec:comparison-grb}, an associated LGRB cannot be ruled out, so a more precise term would be ``apparently orphan.''
We use ``on-axis'' to refer to a viewing angle that is within the opening angle of the initial relativistic material.

We assume a flat $\Lambda$CDM cosmology with H$_0=67.7\,\km\,\psec\,\pmpc$ and $\Omega_M=0.307$ \citep{Planck2016}.
Times are presented in UTC, and magnitudes are given in AB \citep{Oke1983}.
The optical photometry and spectroscopy will be made public through WISeREP, the Weizmann Interactive Supernova Data Repository \citep{Yaron2012}.

\section{Observations} \label{sec:observations}

\subsection{ZTF}

The ZTF custom mosaic camera \citep{Dekany2020} is mounted on the 48-inch Samuel Oschin Telescope (P48) at Palomar Observatory.
As summarized by \citet{Bellm2019_surveys}, during Phase I (ZTF-I; \edit1{2018--2020})
observing time was divided between public (40\%), partnership (40\%), and Caltech (20\%) surveys.
During ZTF-II it is 50\% public and 30\% partnership.
Three custom filters are used ($g_{\mathrm{ZTF}}$, $r_{\mathrm{ZTF}}$, and $i_{\mathrm{ZTF}}$; hereafter $g$, $r$, and $i$, respectively; \citealt{Dekany2020})
and images reach a typical dark-time limiting magnitude of $r\approx20.5\,$mag.

Images are processed and reference-subtracted
by the IPAC ZTF pipeline \citep{Masci2019} using the \citet{Zackay2016} image-subtraction algorithm.
Every 5$\sigma$ point-source detection is saved as an ``alert.''
Alerts are distributed in Avro format \citep{Patterson2019} and can be filtered based on a machine-learning real-bogus metric \citep{Mahabal2019,Duev2019}; host-galaxy characteristics, including a star-galaxy classifier \citep{Tachibana2018}; and light-curve properties.
During ZTF-I the collaboration used a web-based system called the GROWTH marshal \citep{Kasliwal2019} to identify, monitor, and coordinate follow-up observations for transients of interest.
In ZTF Phase II (ZTF-II; 2020--present) the collaboration uses the Fritz marshal \citep{vanderWalt2019,Duev2019}.

\edit1{The ten afterglows discovered to date by ZTF were identified by several different surveys.}
Four events were discovered as part of high-cadence (HC) observations --- either the high-cadence partnership survey, which covered 2500\,deg$^{2}$ with six visits per night (three in $r$ band and three in $g$ band), or the ZTF Uniform Depth Survey (ZUDS\footnote{\url{https://github.com/zuds-survey/zuds-pipeline}}), which covered 2500\,deg$^{2}$ with six visits per night (2$r$, 2$g$, and 2$i$).
Three events were discovered in $gr$ one-day cadence data, including two from public observations shadowing the Transiting Exoplanet Survey Satellite (TESS; \citealt{Ricker2014}) fields \citep{TESS2019}.
Finally, two events were discovered as part of the public ZTF-II all-sky survey, which covers 15,000\,\degsq\ in $r$ band and $g$ band every two nights.
\edit1{An additional afterglow (AT\,2019pim; Perley et al., in prep.) was identified in follow-up observations to a gravitational-wave trigger \citep{Kasliwal2020}.}

\subsection{Search Procedure}
\label{sec:search-procedure}

Every night, the ZTF alert stream is filtered by several independent pipelines to identify young or fast-evolving transients.
In this paper we focus on events discovered via the approach described by \citet{Ho2020d} and \citet{Perley2021}.
In short, basic cuts are applied to remove artifacts, asteroids, and stellar flares.
Remaining transients are divided into several groups, including new transients (those with no detections prior to the current night).
One of us (A.Y.Q.H., D.A.P., Y.Y.) visually inspects the new transients and determines whether any meet the following criteria for afterglows.

\begin{enumerate}
    \item A fast rise from the previous nondetection ($\gtrsim 0.5\,$mag\,\pday).
    \item Red colors ($g-r>0\,$mag) expected from optically thin synchrotron radiation (see \citealt{Ho2020d}) {\it or} rapid intranight fading in a single band.
    \item Either no, or a very faint, associated host in deep archival imaging from the Legacy Survey \citep{Dey2019} or Pan-STARRS1 \citep{Chambers2016}.
\end{enumerate}

\edit1{The criteria listed above were designed to identify on-axis afterglows with high confidence, because our scientific focus is on dirty fireballs.
Off-axis afterglows might rise more slowly and therefore not pass our cuts \citep{vanEerten2010}.
We defer a discussion of search criteria for off-axis afterglows to future work.
In addition, we expect to miss afterglows that are first detected late in their evolution; this is better suited to filters based solely on transient decay rate, such as ZTFReST \citep{Andreoni2020,Andreoni2021}.}

Our search criteria evolved over time.
Based on a strategy developed to discover afterglows in iPTF data \citep{Ho2018}, we initially searched for afterglows via rapid intranight fading.
However, the intranight-fading approach has two limitations: it requires multiple observations per night in a single filter, and ZTF only obtains high-cadence observations across a few thousand square degrees of sky. In addition, when an afterglow is discovered it has already faded significantly, making spectroscopic follow-up observations more difficult.
To discover afterglows earlier in their evolution and over a wider area of sky, we broadened our strategy to also include events that rapidly brighten \citep{Ho2020d}.
This approach enabled us to discover two events in the two-day cadence all-sky public survey (AT\,2021cwd and AT\,2021lfa) that showed no significant fading in the ZTF data:
in fact, each event had only one $g$-band and one $r$-band measurement in the alert stream.
We discuss the possibilities enabled by discovering afterglows in the all-sky survey in \S\ref{sec:conclusions}.

\edit1{After} a candidate is identified by the daily scanner, we check for associated GRBs, obtain follow-up observations to confirm the afterglow nature, and publicly announce the discovery of the transient.
In most cases, we obtain deep imaging to measure the rate of fading and obtain a better constraint on the color.
We primarily use the Liverpool Telescope (LT; \citealt{Steele2004}) owing to its sensitivity, robotic scheduling, and multiband capabilities. When LT is unavailable, we request observations with the Spectral Energy Distribution Machine (SEDM; \citealt{Blagorodnova2018,Rigault2019}) on the Palomar 60-inch telescope (P60; \citealt{Cenko2006}) or the Growth-India Telescope (GIT).
The latency from the first ZTF detection to the first epoch of follow-up imaging has ranged from 0.2\,d to 0.9\,d.
For identifying an associated detected GRB, 
we search the archives of the third Interplanetary Network (IPN\footnote{\url{http://www.ssl.berkeley.edu/ipn3/index.html}}), which
consists of ten spacecraft that provide all-sky
full-time monitoring for high-energy bursts.
The most sensitive GRB detectors in the IPN are the
Burst Alert Telescope (BAT; \citealt{Barthelmy2005}) onboard the Neil Gehrels \emph{Swift} Observatory (\emph{Swift}; \citealt{Gehrels2004}),
the Gamma-ray Burst Monitor (GBM; \citealt{Meegan2009}) on the \emph{Fermi} spacecraft, and the
KONUS instrument on the \emph{Wind} spacecraft (Konus-\emph{Wind}; \citealt{Aptekar1995}).
If a transient is confirmed to be rapidly fading, and has colors consistent with optically thin synchrotron radiation, we trigger X-ray observations with \emph{Swift} and optical spectroscopy of the transient emission to measure its redshift.
If a candidate is confirmed to be cosmological and is ``orphan'' (no GRB is found post facto), we trigger radio observations with the Very Large Array (VLA; \citealt{Perley2011}).
The time from first ZTF detection to public announcement is typically 24\,hr.

\subsection{Overview of ZTF Afterglows}

\edit1{The search procedure outlined in Section~\ref{sec:search-procedure} has resulted in the discovery of seven afterglows to date.
In this paper we present Keck imaging at the position of the previously published afterglow AT\,2020blt \citep{Ho2020b},
as well as discovery and follow-up details for the six afterglows discovered after AT\,2020blt.
Three additional afterglows have been discovered by ZTF (\citealt{Andreoni2020,Andreoni2021}, Perley et al in prep.) and we do not present any new data on these objects in this paper.
}

\edit1{Table~\ref{tab:summary} summarizes all fast optical transients discovered by optical transient surveys that were classified as afterglows based either on the post-facto detection of a likely GRB counterpart, or confirmation of relativistic ejecta from a redshift measurement or radio observations.}
Pre-ZTF, three afterglows had been discovered using optical surveys: PTF11agg \citep{Cenko2013}, iPTF14yb \citep{Cenko2015}, and ATLAS17aeu \citep{Stalder2017,Bhalerao2017}.
With ZTF, the number of optically discovered afterglows has increased from three (of which two had redshift measurements) to 13 (of which nine have redshift measurements).

\begin{deluxetable*}{lrrrrrrrr} 
\tablecaption{Summary of cosmological fast transients discovered by            optical surveys to date.  \label{tab:summary}} 
\tablewidth{0pt} 
\tablehead{ \colhead{Name} & \colhead{R.A.} & \colhead{Decl.} & \colhead{IAU Name} & \colhead{Disc. Date} & \colhead{Disc. Mag.} & \colhead{Redshift} & \colhead{GRB} & \colhead{Ref.} \\ \colhead{} & \colhead{(J2000)} & \colhead{(J2000)} & \colhead{} & \colhead{(MJD)} & \colhead{(AB)} & \colhead{} & \colhead{} & \colhead{}} 
\tabletypesize{\scriptsize} 
\startdata 
PTF11agg & 08:22:17.195 & $+$21:37:38.26 & -- & 55591.2203 & $R=18.26\pm0.05$ & $0.5 \lesssim z \lesssim 3.0$ & -- & [1] \\
iPTF14yb & 14:45:58.01 & $+$14:59:35.1 & -- & 56714.4289 & $r'=18.16\pm0.03$ & 1.9733 & 140226A & [2] \\
ATLAS17aeu & 09:13:13.89 & $+$61:05:33.6 & -- & 57758.4130 & $i_{P1}=17.75\pm0.01$ & $1 \lesssim z \lesssim 2.9$ & 170105A & [3,4,5] \\
ZTF19abvizsw & 18:37:53.48 & $+$61:29:52.7 & AT\,2019pim & 58728.1799 & $g=20.04\pm0.16$ & 1.2596 & -- & [6,7] \\
ZTF20aajnksq & 12:47:04.87 & $+$45:12:02.3 & AT\,2020blt & 58876.2801 & $r=19.57\pm0.14$ & 2.9 & -- & this work$^\dag$; [8] \\
ZTF20abbiixp & 14:12:10.33 & $+$60:54:19.0 & AT\,2020kym & 58993.2863 & $r=17.35\pm0.04$ & 1.256 & 200524A & this work; [9,10] \\
ZTF20abtxwfx & 16:41:21.21 & $+$57:08:20.5 & AT\,2020sev & 59079.2220 & $r=19.27\pm0.10$ & Unknown & 200817A & [10] \\
ZTF20acozryr & 02:48:44.31 & $+$12:08:14.1 & AT\,2020yxz & 59157.3661 & $g=19.47\pm0.19$ & 1.105 & 201103B & [10] \\
ZTF21aaeyldq & 08:15:15.33 & $-$05:52:01.3 & AT\,2021any & 59230.2916 & $r=17.92\pm0.06$ & 2.5131 & -- & this work; [11] \\
ZTF21aagwbjr & 07:48:19.32 & $+$11:24:34.1 & AT\,2021buv & 59249.2966 & $r=17.11\pm0.05$ & 0.876 & 210204A & this work; [10] \\
ZTF21aakruew & 10:24:42.15 & $+$11:36:40.9 & AT\,2021cwd & 59257.3697 & $g=19.57\pm0.21$ & Unknown & 210212B & this work; [12] \\
ZTF21aayokph & 12:32:48.72 & $-$01:29:22.5 & AT\,2021lfa & 59338.2325 & $r=18.60\pm0.08$ & 1.0624 & -- & this work; [13] \\
ZTF21abfmpwn & 16:15:40.38 & $+$14:23:56.5 & AT\,2021qbd & 59376.2325 & $g=18.49\pm0.10$ & 1.1345 & 210610B & this work; [14] \\
\enddata 
\tablerefs{[1] \citet{Cenko2013}, [2] \citet{Cenko2015}, [3] \citet{Stalder2017}, [4] \citet{Bhalerao2017}, [5] \citet{Melandri2019}, [6] \citet{Perley2019_ZTF19abvizsw}, [7] \citet{Kasliwal2020}, [8] \citet{Ho2020d}, [9] \citet{Ho2020_ZTF20abbiixp}, [10] \citet{Andreoni2021}, [11] \citet{Ho2021GCN}, [12] \citet{Yao2021}, [13] \citet{Yao2021_ZTF21aayokph}, [14] \citet{Perley2021_ZTF21abfmpwn}}
\tablenotetext{\dag}{We provide discovery and/or follow-up data as part of this paper.}
\end{deluxetable*} 

\begin{figure}[hbt!]
    \centering
    \includegraphics[width=\columnwidth]{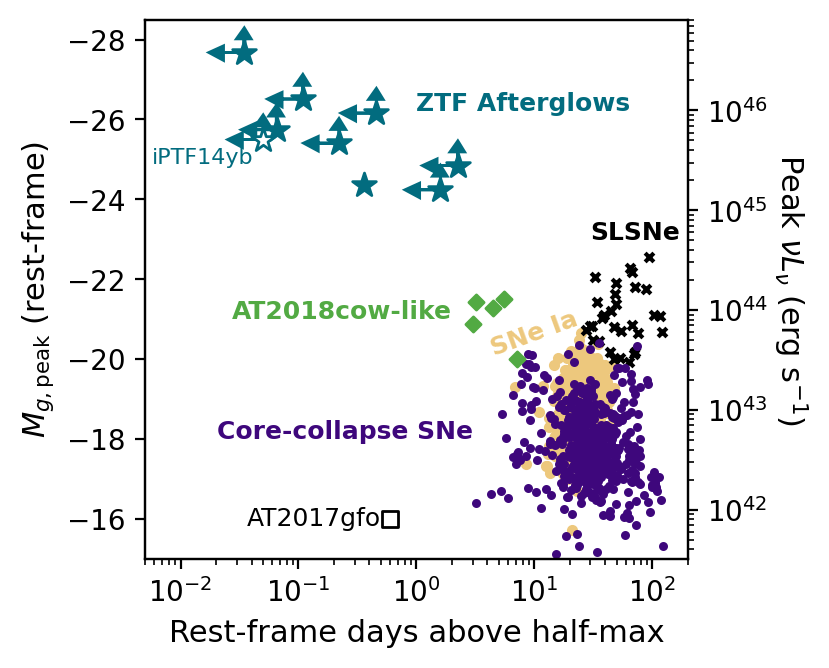}
    \caption{The duration and luminosity of optical transients. Measurements of superluminous supernovae (SLSNe), Type Ia supernovae (SNe~Ia), and most core-collapse supernovae (CC~SNe) are from the ZTF Bright Transient Survey \citep{Fremling2020_RCF,Perley2020_BTS}.
    Measurements of short-duration CC~SNe and AT\,2018cow-like transients are from dedicated searches for fast-evolving transients \citep{Prentice2018,Perley2019cow,Perley2021,Ho2020b,Ho2021_RET,Yao2021_AT2020mrf}.
    For reference we also show the timescale and luminosity of the optical emission from GW170817/AT\,2017gfo \citep{Coulter2017,Cowperthwaite2017,Kasliwal2017,Drout2017,Villar2017_KN} and of iPTF14yb (the only optically discovered afterglow prior to ZTF with a redshift measurement; \citealt{Cenko2015}). The afterglows discussed in this paper are the fastest and most luminous optical transients discovered and monitored in real time. Owing to the cadence of ZTF, our measurements typically represent upper limits on the duration and lower limits on the peak luminosity. Measurements are in the rest-frame $g$ band when possible, with a crude $K$-correction applied (Equation~\ref{eq:opt-lc} for afterglows; see \citet{Ho2021_RET} for other sources).}
    \label{fig:lum-time}
\end{figure}

The afterglows in Table~\ref{tab:summary} constitute the shortest-lived optical extragalactic transients that have been discovered in optical survey data and followed up in real time.
To illustrate this, Figure~\ref{fig:lum-time} shows the duration above half-maximum intensity and the peak absolute magnitude for optical transients, primarily with light curves observed by ZTF \citep{Fremling2020_RCF,Perley2020_BTS,Ho2021_RET}.
Most supernovae (SNe) evolve on timescales from \hbox{10--100}\,d, powered by radioactive decay, with their characteristic duration set by diffusion through optically thick ejecta \citep{Villar2017}.
By contrast, the afterglow emission is governed by optically thin synchrotron radiation.
We caution that for the afterglows, estimates of the duration and peak luminosity are imprecise because the ZTF cadence is much slower than the light-curve timescale; the exception is AT\,2019pim owing to TESS observations of the light curve (Perley et al., in prep; \citealt{Fausnaugh2019}).
To estimate the duration, we use best-fit power laws to the light curve (\S\ref{sec:comparison-grb}).
The luminosity estimates are described in \S\ref{sec:comparison-opt-lc}.

In the remainder of this section we provide discovery and follow-up details for events discovered by our search procedure.
The X-ray, optical, and radio \edit1{light curves} are respectively provided in Tables~\ref{tab:opt-phot}, \ref{tab:xray-phot}, and \ref{tab:radio-phot} in the Appendix.
When appropriate, we estimate the chance spatial and temporal coincidence of the optical transient with a GRB, by calculating the number of LGRBs we expect a given facility to detect in the localization region during the time interval of interest.
For \emph{Fermi}-GBM, we use the fact that during the year 2020 GBM detected 260 bursts, for a rate of 0.7\,\pday.
For the IPN, we use the fact that according to the IPN master list\footnote{http://www.ssl.berkeley.edu/ipn3/masterli.txt}, during the year 2020 IPN detected 422 bursts, an average of 1.5\,d$^{-1}$.

\subsubsection{ZTF\,20abbiixp / AT\,2020kym / GRB\,200524A}
\label{sec:obs-ZTF20abbiixp}

ZTF\,20abbiixp was first detected by ZTF on 2020 May 24.29 (MJD 58993.29) in an $r$-band image at $r=17.35\pm0.04\,$mag as part of the ZTF Uniform Depth Survey.
The most recent nondetection was 0.84\,d prior at $g>20.51\,$mag.
The most recent $r$-band nondetection was 1.01\,d prior at $r>20.79\,$mag, giving an $r$-band rise rate of $>3.4\,$mag\,\pday.

\begin{figure*}[hbt!]
    \centering
    \includegraphics[width=0.8\textwidth]{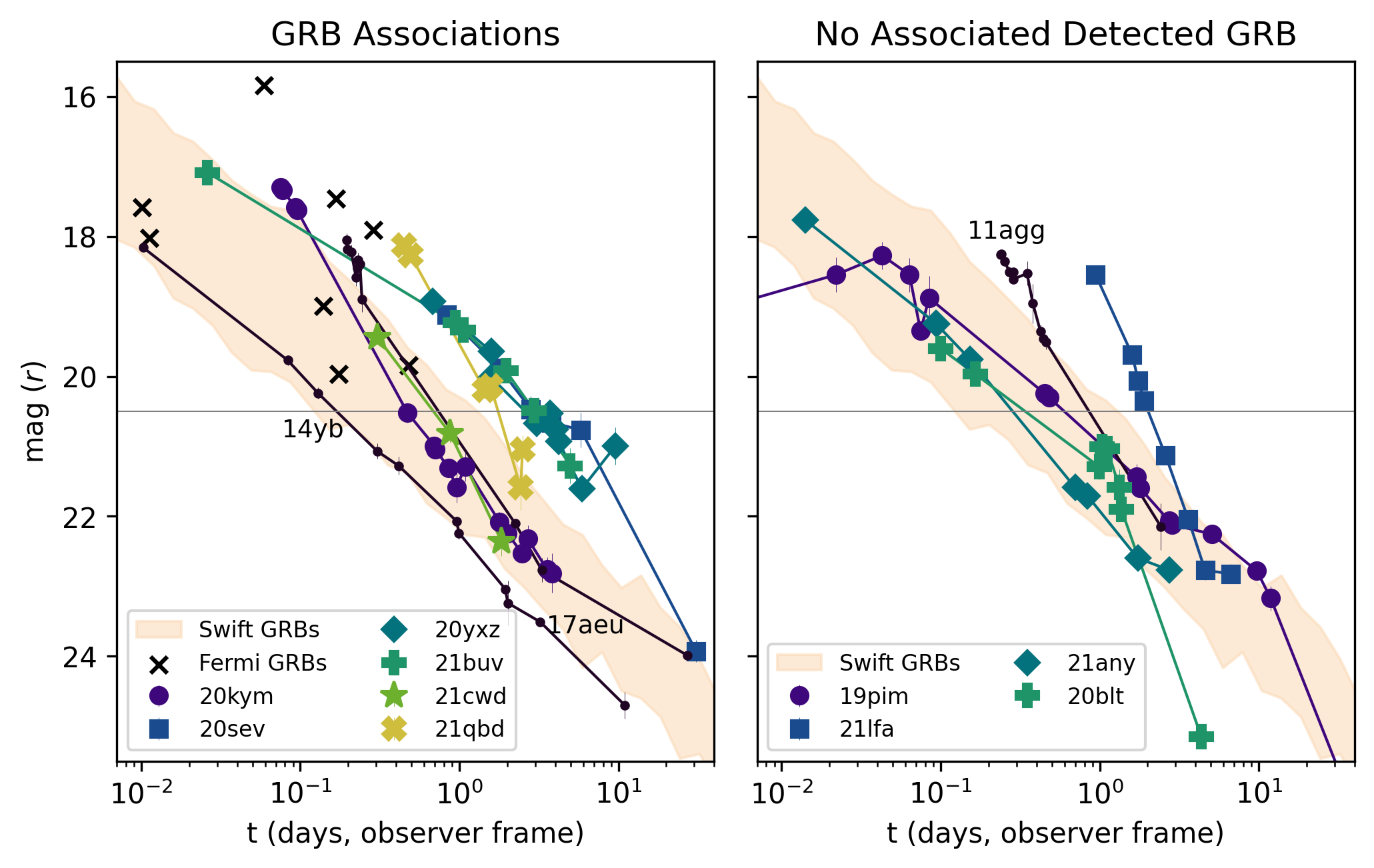}
    \caption{The $r$-band light curve (corrected for Milky Way extinction) of each ZTF afterglow compared to the sample of afterglows detected in follow-up observations to \emph{Swift}-BAT triggers \citep{Kann2010} and \emph{Fermi}-GBM triggers \citep{Singer2015}.
    The left panel shows afterglows with associated detected GRBs, and the right panel indicates afterglows with no associated detected GRB.
    For the ``orphan'' events, we caution that the estimated $t_0$ is uncertain.
    The shaded region corresponds to the 25 to 75 percentile bounds of the \emph{Swift}-BAT sample \citep{Kann2010}.
    The crosses indicate the first detection of each \emph{Fermi}-GBM afterglow in \citet{Singer2015}.
    The horizontal gray line indicates the nominal ZTF limit of 20.5\,mag.
    Photometry is obtained from \citet{Cenko2013,Cenko2015}, \citet{Bhalerao2017}, \citet{Ho2020d}, \citet{Andreoni2021}, and Perley et al. (in prep.).
    For ATLAS17aeu we include $c$-band points. For AT\,2019pim we include TESS points obtained at $\Delta t<0.1\,$d.
    Our searches tend to find afterglows that are brighter than the bulk of the follow-up sample.}
    \label{fig:compare-opt-lc-kann}
\end{figure*}

There were five detections the first night, all in $r$\edit1{ band}, which showed significant fading of 0.35\,mag over 0.48\,hr \edit1{(18\,mag\,\pday)};
we provide the full set of ZTF photometry in Table~\ref{tab:opt-phot},
and plot the $r$\edit1{-band} light curve in Figure~\ref{fig:compare-opt-lc-kann}.
Legacy Survey \edit1{pre-}imaging \citep{Dey2019} showed no associated host at the transient position down to $g\approx r\approx24\,$mag.
The fast rise, rapid intranight fading, and lack of detected host in deep imaging led the transient to be flagged by the daily scanner on 2020 May 24.8.
ZTF\,20abbiixp also passed the ZTFReST search criteria during their science validation \citep{Andreoni2021}.

A \edit1{post facto} search for an associated GRB identified the long-duration GRB\,200524A \citep{Pookalil2020_200524A,Dirirsa2020_200524A} consistent with the position and time of ZTF\,20abbiixp.
GRB\,200524A had triggered the \emph{Fermi}-GBM and the \emph{Fermi}-Large Area Telescope (LAT; \citealt{Atwood2009}),
and \emph{Swift} X-ray Telescope (XRT; \citealt{Burrows2005}) ToO observations had been initiated \citep{Evans2020_200524A}.
The burst was found to also have triggered ASTROSAT \citep{Gupta2020_200524A,Astrosat2014} and Konus-\emph{Wind} \citep{Svinkin2020_200524A}.

The trigger time was 1.8\,hr prior to the first ZTF detection. The offset was 0.151\,deg from the LAT position, which had a localization region of radius 0.2\,deg (90\% containment, statistical error only).
The expected number of GRBs detected by GBM in this location over a 0.84\,d window is $R_\mathrm{GBM} \times 0.84\,\days \times (\pi (0.2)^2) / 41253 = 2\times10^{-6}$.
We therefore consider the association secure.
We publicly reported the transient \citep{Ho2020_ZTF20abbiixp} and saved it to the Transient Name Server (TNS\footnote{\url{https://www.wis-tns.org/}}), where it was assigned the name AT\,2020kym. 
Our public report of the ZTF transient and likely GRB association represented the first identification of the afterglow of GRB\,200524A.

GRBs detected by the \emph{Fermi}-LAT are of general interest because they typically have a relativistic energy release that is an order of magnitude or more greater than the canonical $10^{51}\,\erg$ value,
as well as brighter-than-average X-ray and optical afterglows \citep{Nysewander2009,Cenko2011,McBreen2010}.
A multiwavelength analysis of AT\,2020kym will be published in a separate paper by Ghosh et al., so here we simply summarize the observations that we and other groups obtained.

We obtained a long-slit spectrum of AT\,2020kym between $\Delta t = 26.6\,$hr and $\Delta t = 27.8$\,hr with the Gemini Multi-Object Spectrograph (GMOS) on Gemini-North \citep{Hook2004} under our Target-of-Opportunity (ToO) program\footnote{GN-2020A-Q-117; PI: Miller}, and as discussed by \citet{Yao2021_ZTF20abbiixp_redshift} measured a redshift of $z=1.256$. The spectrum will be published by Ghosh et al.

Because of the interest in \emph{Fermi}-LAT GRBs,
a variety of follow-up observations were obtained for this event.
We triggered our VLA program\footnote{Project ID VLA/20A-374; PI A. Ho} beginning 5\,d after the burst and obtained several observations at 10\,GHz, which will be presented by Ghosh et al.
The X-ray afterglow was detected by \emph{Swift}-XRT \citep{Capalbi2020_200524A}.
Optical photometry was obtained with a large number of facilities \citep{Kumar2020_200524A,Lipunov2020_200524A,Rumyantsev2020_200524A,Sanwal2020_200524A,Belkin2020_200524A,Izzo2020_200524A,Perley2020_200524A,deUgartePostigo2020_200524A,Hosokawa2020_200524A,Kumar2020_DFOT_200524A,Zheng2020_200524A,Blazek2020_200524A,Kuin2020_200524A}.

\begin{figure*}[!htb]
    \centering
    \includegraphics[]{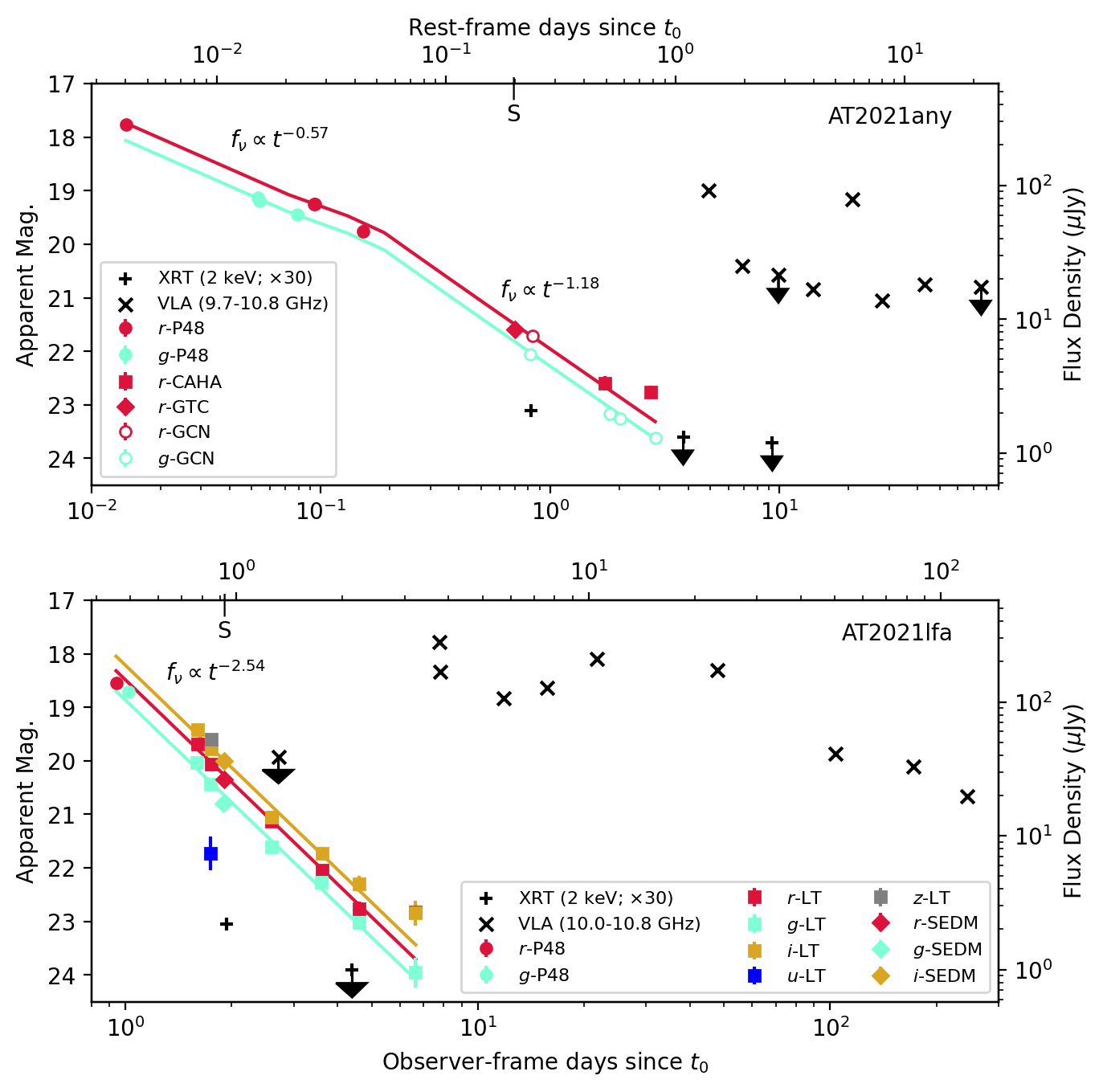}
    \caption{The optical, X-ray, and radio light curves of two new ``orphan'' afterglows, AT\,2021any (top panel) and AT\,2021lfa (bottom panel), together with their best-fit power laws.
    Data presented in this paper are shown as filled points, with different symbols for different instruments.
    Data obtained from GCNs \citep{Zhu2021_ZTF21aaeyldq,Guelbenzu2021_ZTF21aaeyldq} are shown as unfilled circles.
    The optical data have been corrected for Milky Way extinction.
    X-ray data are shown with a plus sign and scaled by a factor of 30 for clarity.
    Radio X-band data are shown with crosses.
    Arrows indicate upper limits.
    An ``S" along the top marks an epoch of spectroscopy.
    The best-fit $t_0$ is calculated as described in \S\ref{sec:comparison-grb} --- we caution that these values are uncertain.
    }
    \label{fig:orphan-lcs}
\end{figure*}

\subsubsection{ZTF\,21aaeyldq / AT\,2021any}

ZTF\,21aaeyldq was first detected by ZTF on 2021 January 16.29 (MJD 59230.29) in an $r$-band image at $r=17.92\pm0.06\,$mag as part of the high-cadence partnership survey.
The most recent nondetection was 20.3\, min prior at $r>20.28\,$mag (in a public-data image), giving a rise rate of $>167\,$mag\,\pday.
There were six detections the first night in $r$ band and $g$ band, which showed significant fading of 1.9\,mag over 3.3\,hr \edit1{(14\,mag\,\pday)} and an extinction-corrected red color of $g-r=(0.25\pm0.19)\,$mag.
The ZTF photometry is presented in Table~\ref{tab:opt-phot}, the $g$- and $r$-band light curves are shown in Figure~\ref{fig:orphan-lcs}, and the $r$-band light curve is shown compared to other events in the right panel of Figure~\ref{fig:compare-opt-lc-kann}.
No host galaxy was visible at the transient position in deep Legacy Survey 
pre-imaging ($>24$\,mag).
Owing to the fast fading, red color, and lack of detected host, the transient was flagged by the daily scanner on 2021 January 16.75.

We searched the \emph{Fermi}-GBM Burst Catalog\footnote{\url{https://heasarc.gsfc.nasa.gov/W3Browse/fermi/fermigbrst.html}},
the \emph{Fermi}-GBM Subthreshold Trigger list\footnote{\url{https://gcn.gsfc.nasa.gov/fermi\_gbm\_subthresh\_archive.html}} (with reliability flag not equal to 2), the
\swift\ GRB Archive\footnote{\url{https://swift.gsfc.nasa.gov/archive/grb\_table/}}, and the Gamma-Ray Coordinates Network archives\footnote{\url{https://gcn.gsfc.nasa.gov/gcn3\_archive.html}} for an associated GRB between the last ZTF nondetection (2021 January 16.28) and the first ZTF detection 20\,min later.
No associated GRB had been reported.
We searched the pointing history of different satellites, and determined that the transient position was visible to \emph{Fermi}-GBM for 14\,min out of the 20\,min in the interval between the last nondetection and the first detection.
The position was not visible to \swift-BAT or to SPI-ACS \citep{Vedrenne2003} onboard the \emph{INTernational Gamma-ray Astrophysics
Laboratory} (\emph{INTEGRAL}; \citealt{Winkler2003}):
\emph{INTEGRAL} was not obtaining data due to its regular perigee passage.
Konus-\emph{Wind} was only recording data in its S2 detector, but despite the high
incident angle to the source (115\,deg), reasonable upper limits on the
$\gamma$-ray emission may be derived (\S\ref{sec:comparison-grb}).
We reported the transient to the TNS on January 16.9 and it was assigned the name AT\,2021any.

  \begin{figure*}[!ht]
   \centering
   \includegraphics[width=0.9\hsize]{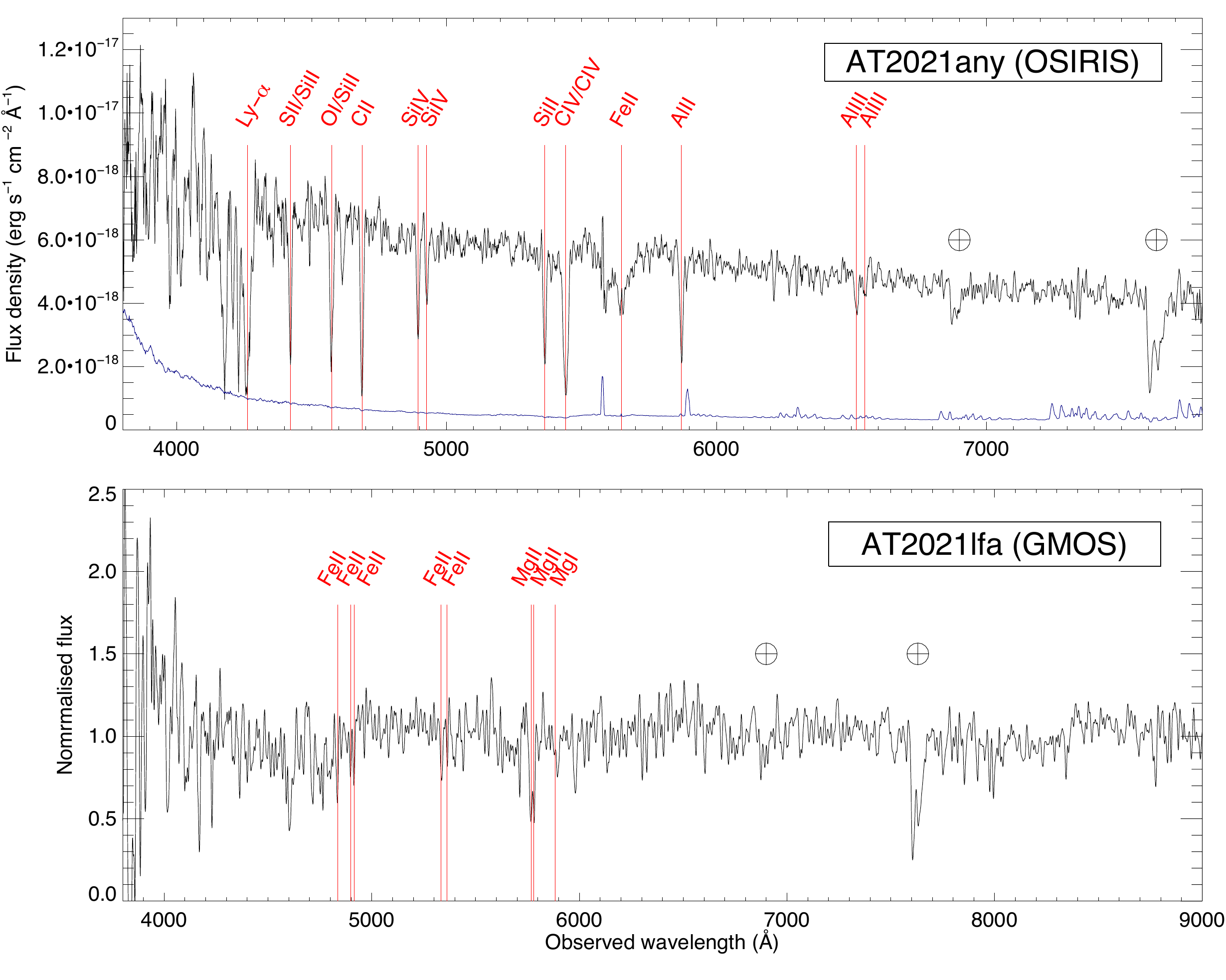}
      \caption{Optical spectra of AT\,2021any (top panel; OSIRIS/GTC) and AT\,2021lfa (bottom panel; GMOS-S/Gemini), two afterglows with no associated detected GRB.
      Vertical lines mark absorption features used to measure the redshifts of $z=2.5131$ (AT\,2021any) and $z=1.0624$ (AT\,2021lfa). Regions affected by telluric absorption are also marked.
      The blue line in the AT\,2021any panel is the error spectrum.
      }
   \label{fig:spec}
   \end{figure*}

AT\,2021any was observed with the Optical Spectrograph and InfraRed Imager System (OSIRIS; \citealt{Cepa2000}) mounted on the 10.4\,m Gran Telescopio Canarias (GTC) telescope at the Roque de los Muchachos Observatory in the island of La Palma (Spain). The observation consisted of $4 \times 900$\,s exposures with grism R1000B and a 1$^{\prime\prime}$ wide slit, aligned with the parallactic angle \citep{Filippenko1982}, which results in a resolving power of $\sim600$ and spectral coverage between 3700\,\AA\ and 7800\,\AA. The observation was performed at a mean epoch of 00:06 UT on 17 January 2021. 
The data reduction was performed using a custom pipeline based on shell scripts and IRAF routines that included bias, response (flatfielding), wavelength calibration, and flux calibration based on the observation of a spectrophotometric star. 
Telluric features were not removed.

As mentioned by \citet{deUgartePostigo2021_ZTF21aaeyldq},
the spectrum (see Figure~\ref{fig:spec}) shows several prominent absorption features corresponding to Ly-$\alpha$, \ion{S}{2}, \ion{O}{1}, \ion{Si}{2}, \ion{Si}{4}, \ion{C}{2}, \ion{C}{4}, \ion{Fe}{2}, \ion{Al}{2}, and \ion{Al}{3} at a common redshift of $z=2.5131\pm0.0016$. To determine the redshift we use the average value calculated for each of the unblended features, taking their standard deviation as the uncertainty. The equivalent widths measured for these lines are displayed in Table~\ref{tab:EW}, together with their measured wavelength and redshift. The reduced spectra can be plotted and downloaded within the GRBSpec database\footnote{\url{http://grbspec.eu}} \citep{deUgartePostigo2014}.

\begin{deluxetable}{lrrrr}[htb!]
\tablecaption{Equivalent widths from optical spectra.
\label{tab:EW}}
\tablewidth{0pt}
\tablehead{\colhead{Name} & \colhead{Obs. $\lambda$} & \colhead{Feature} & \colhead{$z$} & \colhead{EW} \\[-0.3cm]
& \colhead{(\AA)} & \colhead{} & & \colhead{(\AA)}
}
\tabletypesize{\scriptsize} 
\startdata
AT\,2021any & 4421.02 & \ion{S}{2} 1259.52 & 2.5131 & $7.41\pm1.26$ \\
 & & \ion{Si}{2} 1260.42 & 2.5131 &  \\
 &4574.15 & \ion{O}{1} 1302.17 & 2.5131 & $9.31\pm1.03$ \\
 & & \ion{Si}{2} 1304.37 & 2.5131 &  \\
 &4686.88 & \ion{C}{2} 1334.53 & 2.5120 & $10.13\pm0.94$ \\
 &4894.81 & \ion{Si}{4} 1393.76 & 2.5119 & $5.38\pm0.88$ \\
 &4926.22 & \ion{Si}{4} 1402.77 & 2.5118 & 3.46 $\pm$ 0.77 \\
 &5364.17 & \ion{Si}{2} 1526.71 & 2.5135 & 7.81 $\pm$ 0.74 \\
 &5440.73 & \ion{C}{4} 1548.2 & 2.514 & 17.58 $\pm$ 0.98 \\
 & & \ion{C}{4} 1550.77 & 2.5131 &  \\
 &5647.44 & \ion{Fe}{2} 1608.45 & 2.5111 & 8.98 $\pm$ 0.91 \\
 &5870.30 & \ion{Al}{2} 1670.79 & 2.5135 & 7.82 $\pm$ 0.85 \\
 &6518.77 & \ion{Al}{3} 1854.72 & 2.5147 & 2.97 $\pm$ 0.64 \\
 &6549.65 & \ion{Al}{3} 1862.79 & 2.5160 & 1.49 $\pm$ 0.52 \\
AT\,2021lfa & 4834.99 & FeII 2344.21 & 1.0625 & 2.51 $\pm$ 0.82 \\
& 4897.54 & \ion{Fe}{2} 2374.46 & 1.0626 & 2.65 $\pm$ 0.85 \\
& 4914.12 & \ion{Fe}{2} 2382.77 & 1.0624 & 3.52 $\pm$ 0.90 \\
& 5334.21 & \ion{Fe}{2} 2586.65 & 1.0622 & 3.12 $\pm$ 0.95 \\
& 5363.12 & \ion{Fe}{2} 2600.17 & 1.0626 & 3.94 $\pm$ 0.92 \\
& 5768.26 & \ion{Mg}{2} 2796.30 & 1.0628 & 3.30 $\pm$ 0.85 \\
& 5780.84 & \ion{Mg}{2} 2803.50 & 1.0620 & 4.63 $\pm$ 0.88 \\
\enddata
\end{deluxetable}

Preceding our spectroscopic observations with OSIRIS, we obtained two acquisition images of 60\,s and 30\,s in the $r^\prime$-band filter (first reported by
\citealt{deUgartePostigo2021_ZTF21aaeyldq}). The afterglow is clearly detected in each
image. No calibration frames were taken for these observations, but the
raw images are of high quality.

After a weather-induced delay, we obtained two epochs of observations
with the Calar Alto Faint Object Spectrograph (CAFOS) mounted on the
2.2\,m telescope of the Centro Astron\'omico Hispano en Andaluc\'ia
(CAHA), Almeria, Spain, in the $R_C$ filter (first reported by
\citealt{Kann2021_ZTF21aaeyldq}, and \citealt{Kann2021_ZTF21aaeyldq_jet_break}). The observations consisted
of $10\times360$\,s at $\sim1.7$\,d after the GRB, and another
$12\times360$\,s a night later. Observing conditions were fair, but with
mediocre seeing. The afterglow is clearly detected in each stack. Images
were bias-subtracted, flatfielded, aligned, and stacked with standard
procedures in ESO
MIDAS\footnote{\url{https://www.eso.org/sci/software/esomidas/}} and
PyRAF\footnote{\url{https://iraf-community.github.io/pyraf.html}}.
The CAFOS and OSIRIS photometry is presented in Table~\ref{tab:opt-phot}.
Additional optical follow-up observations of AT\,2021any were reported by a variety of facilities \citep{Kumar2021_ZTF21aaeyldq,Ahumada2021_ZTF21aaeyldq,Zhu2021_ZTF21aaeyldq,Hu2021_ZTF21aaeyldq,Coughlin2021_ZTF21aaeyldq,Rossi2021_ZTF21aaeyldq,Guelbenzu2021_ZTF21aaeyldq,Ghosh2021_ZTF21aaeyldq}.

We obtained a \emph{Swift}-XRT\footnote{PI A. Ho; target ID 13991} observation of AT\,2021any beginning 0.81\,d after the first optical detection.
We obtained three epochs
of 3\,ks exposures and reduced the data using the online tool\footnote{\url{https://www.swift.ac.uk/user objects/}} developed by the \emph{Swift} team \citep{Evans2007}. In the first epoch X-ray emission was detected at the transient position, and there was no detection in the subsequent two epochs.
The observation log is provided in Table~\ref{tab:xray-phot}.
Taking a neutral hydrogen column density of $n_H = 8.12 \times10^{20}\,\pcmsq$ \citep{Willingale2013} and assuming a power-law spectrum with photon index $\Gamma=2$, we find an unabsorbed flux density of $3.30\times10^{-13}\,\erg\,\pcmsq\,\psec$ using
\texttt{webpimms}\footnote{\url{https://heasarc.gsfc.nasa.gov/cgi-bin/Tools/w3pimms/w3pimms.pl}}.
The X-ray light curve is shown in Figure~\ref{fig:orphan-lcs}.

\begin{figure*}[hbt!]
    \centering
    \includegraphics[width=0.9\textwidth]{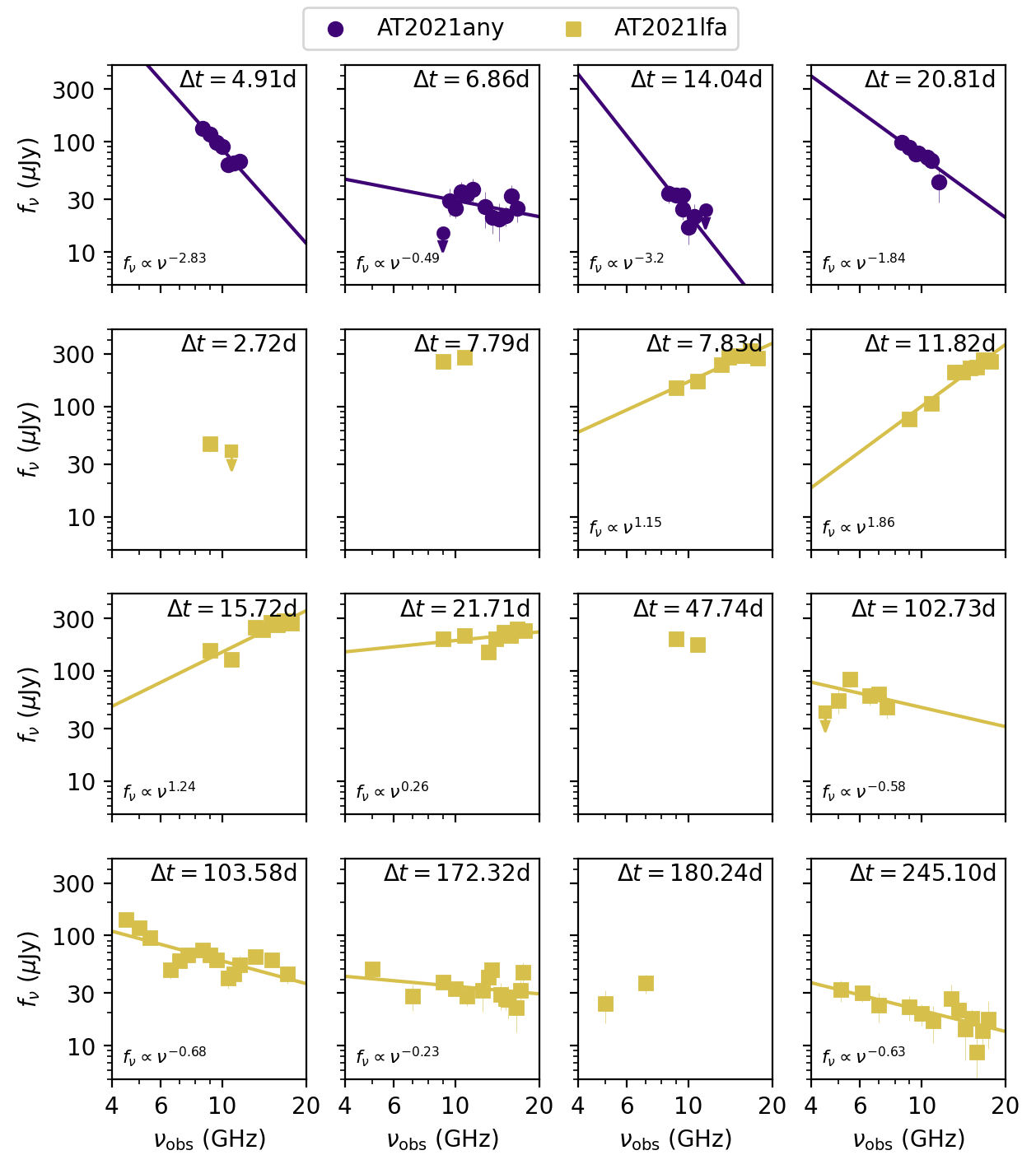}
    \caption{Evolution of the radio SED of AT\,2021any (circles) and AT\,2021lfa (squares), afterglows with no associated detected GRB. The lines show power-law fits to the data for each epoch with $>2$ data points.
    The SEDs of AT\,2021any appear optically thin throughout.
    The SEDs of AT\,2021lfa appear self-absorbed up to 21.71\,d and then become optically thin, suggesting that the self-absorption frequency has passed through the VLA observing bands.
    Note that the light curve of AT\,2021lfa shows evidence for scintillation at $\nu\lesssim10\,$GHz.
    }
    \label{fig:radio-sed}
\end{figure*}

We obtained eight epochs of VLA observations of AT\,2021any under our ToO program\footnote{VLA/20B-164 and VLA/21A-319, PI D. Perley}: six epochs were in X-band (10\,GHz), one epoch was in X- and Ku-band (15\,GHz), and one epoch was in S- (3\,GHz), C- (6\,GHz), X-, and Ku-band.
The source J0808-0751 was used as a phase calibrator.  For most observations 3C138 was used as the flux calibrator, although to account for the recent flaring behavior of this source we also observed 3C286 during the third epoch, and in the final two epochs we observed 3C286 only.
Reduction of the data was performed using the Astronomical Image Processing System (\texttt{AIPS}) using standard synthesis imaging techniques.  Calibration was performed by hand, with regions of the spectrum heavily contaminated by radio-frequency interference (RFI) excluded.
We \edit1{publicly} reported the first detection \citep{Perley2021_ZTF21aaeyldq} and the observations are listed in Table~\ref{tab:radio-phot}.  The 10\,GHz radio light curve is shown in Figure~\ref{fig:orphan-lcs},
and the evolution of the radio spectral energy distribution (SED) is shown in Figure~\ref{fig:radio-sed}.

\edit1{We obtained a deep image of the position of AT\,2021any using the Low Resolution Imaging Spectrometer (LRIS; \citealt{Oke1995}) on the Keck-I 10\,m telescope. We imaged in four filters ($U$, $G$, $R$, and $RG850$) with exposure times of 20\,min per filter. Data reduction was performed using LPipe \citep{Perley2019lpipe}. The $U$, $G$, and $RG850$ filters were calibrated relative to the SDSS $u$, $g$, and $z$ bands and reported in the AB system.  Cousins $R$-band magnitudes for secondary standards were calculated via the Lupton transform and magnitudes are reported in the AB system. A faint ($g=25$\,mag) source was detected at the transient position, and the photometry is presented in Table~\ref{tab:host}.
The host colors resemble those of a typical LGRB host galaxy.
}

\begin{deluxetable*}{lrrrrrr}[!htb]
\tablecaption{LRIS imaging of transient locations to search for host counterparts. \label{tab:host}} 
\tablewidth{0pt} 
\tablehead{ \colhead{Name} & \colhead{Date} & \colhead{Exposure} & \colhead{$U$} & \colhead{$G$} & \colhead{$R$} & \colhead{$RG850$} \\[-0.3cm] \colhead{} & \colhead{(UT)} & \colhead{(s)} & \colhead{(mag$^\dag$)} & \colhead{(mag)} & \colhead{(mag)} & \colhead{(mag)}} 
\tabletypesize{\scriptsize} 
\startdata 
AT\,2021any & 2022 January 31 & 1380 & $>$25.90 & -- & -- & $24.99 \pm 0.29$\\
-- & -- & 1080 & -- & $25.02 \pm 0.10$ & $25.27 \pm 0.17$ & -- \\
AT\,2020blt$^\ddag$ & 2022 January 31 & 960 & -- & $>$ 26.49 & $>$ 25.92 & -- \\
-- & 2022 February 27 &1920 &  $> 26.65$ & -- & -- & $>25.34$ \\
AT\,2021lfa & 2022 March 03 & 1080 & $>$25.55 & -- & -- & $>$24.68 \\
-- & -- & 900 & -- & $>$26.50 &  $>26.17$ & -- \\
\enddata
\tablenotetext{\dag}{Magnitudes are reported in the AB system.}
\tablenotetext{\ddag}{For the $RG850$ observation of AT\,2020blt, we created custom sky flats for each exposure using dithered images of the field to correct for variations in the flat-field pattern between the science and calibration frame.}
\end{deluxetable*}

\subsubsection{ZTF\,21aagwbjr / AT\,2021buv / GRB\,210204A}

ZTF\,21aagwbjr was first detected in an image obtained on 2021 February 04.30 at $r=17.11\pm0.05$\,mag as part of the one-day cadence TESS shadowing survey. The previous nondetection was 1.87\,hr prior at $g>18.25$\,mag. The previous $r$-band nondetection was 1.05\,d prior at $r>20.35\,$mag, giving a rise rate in $r$ band of $>3.1\,$mag\,\pday.
There was only one detection the first night.
There were four detections the second night, two in $r$ band and two in $g$ band.
Owing to the rapid fade rate of $2.15\pm0.14\,$mag\,\pday\ in $r$ band, the extinction-corrected red color ($g-r=0.42\pm0.19\,$mag), and the presence of a faint ($g=24.6\,$mag, $r=23.7\,$mag) counterpart (putative host galaxy) at the transient position in Legacy Survey \edit1{pre-}imaging,
the transient was flagged by the daily scanner on February 05.7.
ZTF\,21aagwbjr was also flagged by the ZTFReST pipeline, and the ZTF photometry was presented by \citet{Andreoni2021}.

A search for an associated GRB identified the long-duration GRB\,210204A consistent with the position and time of ZTF\,21aagwbjr.
GRB\,210204A had triggered \emph{Fermi}-GBM \citep{GBM_210204A}, the Gamma-Ray Detector (GRD) onboard the \emph{Gravitational Wave High-energy Electromagnetic Counterpart All-sky Monitor} (\emph{GECAM}; \citealt{GECAM2019,GECAM_210204A}),
Konus-\emph{Wind} \citep{Frederiks2021,Frederiks2021_corr}, and \emph{AstroSat} \citep{Waratkar2021}.
The trigger time of February 04.27 was 43\,min prior to the first ZTF detection.
The IPN localized the burst to a region of 6.7\,\degsq\ \citep{Hurley2021}.
The number of expected GRBs detected by the IPN in this region over the 1.87\,hr window is $2\times10^{-5}$, so the association is quite secure.
The transient was reported \citep{Kool2021} and saved to the TNS, where it was assigned the name AT\,2021buv.
The multiwavelength properties of AT\,2021buv will be published in a separate paper by Kumar et al. Here we summarize the follow-up observations that were obtained.

We measured the redshift of AT\,2021buv using
a long-slit spectrum obtained with GMOS-S\footnote{ToO program GS-2021A-Q-124; PI A. Ho} \citep{Gimeno2016}. The observation was conducted in the Nod-and-Shuffle mode with a 1\farcs0-wide slit, beginning 42.8\,hr after the \emph{Fermi}-GBM trigger. We obtained $2 \times 450$\,s spectroscopic exposures with the B600 grating and $2 \times 450$\,s exposures with the R400 grating, providing coverage over the range 3620--9600\,\AA. We reduced the spectrum using the \texttt{IRAF} package for GMOS, and clearly detected \ion{Mg}{2} and \ion{Mg}{1} in absorption at $z=0.876$.
The redshift of 0.876 was independently measured using VLT/X-shooter \citep{Xu2021} from absorption features including fine-structure lines of \ion{Fe}{2}, and emission lines of \ion{O}{2}, \ion{O}{3}, H$\beta$, and H$\alpha$.
The GMOS spectrum will be published by Kumar et al.

AT\,2021buv was also detected in the X-ray \citep{Swift2021_ZTF21aagwbjr,Kennea2021} and radio \citep{Chandra2021_ZTF21aagwbjr} bands.
Additional optical photometry has been made available \citep{Teja2021,Belkin2021_ZTF21aagwbjr,Rossi2021,Gupta2021_GRB210204A} and includes reports of a jet break \citep{Gupta2021_GRB210204A,Rossi2021}.

\subsubsection{ZTF\,21aakruew / AT\,2021cwd / GRB\,210212B}

ZTF\,21aakruew was first detected in a ZTF image obtained on 2021 February 12.37 at $g=19.57\pm0.21\,$mag, as part of the public survey.
The last nondetection was 2.0\,d prior at $g>20.88\,$mag, giving a rise rate of $>0.7\,$mag\,\pday.
The extinction-corrected color was $g-r=0.02\pm0.26\,$mag and there was no host galaxy visible at the transient position in Legacy Survey \edit1{pre-}imaging (down to $g\approx r \approx 24\,$mag).
Owing to the fast rise and lack of detected host, the transient was noted by the daily scanner on February 12.7.

The uncertainty in the $g-r$ color was not sufficient to determine whether it was red as expected for optically thin synchrotron emission ($g-r=0.24\,$mag) or blue as expected for thermal flares ($g-r=-0.17\,$mag).
To determine the color more precisely, we obtained LT $griz$ imaging on February 12.98.
The LT imaging revealed rapid fading of 2.4\,mag\,\pday\ in $r$ band and red colors (extinction-corrected $g-r=0.75\pm0.21\,$mag).
We report our optical photometry in Table~\ref{tab:opt-phot}.

We could not identify any GRB consistent with the position and time of ZTF\,21aakruew.
We publicly announced the transient \citep{Yao2021_ZTF21aakruew} and saved it to the TNS, where it was assigned the name AT\,2021cwd.
After our announcement, the IPN announced that AT\,2021cwd was in the localization region of long-duration GRB\,210212B by Konus-\emph{Wind}, \emph{INTEGRAL}, and \emph{Swift}-BAT \citep{Svinkin2021_GRB210212B,Svinkin2021_GRB210212B_2}.
The burst time was 6.2\,hr prior to the first ZTF detection.
Follow-up optical photometry was obtained by us with LT as well as by other groups \citep{Pozanenko2021,Guelbenzu2021}.
A conservative estimate of the 3$\sigma$ IPN localization area is 1286\,\degsq\ (the BAT coded region cannot be confidently excluded).
The number of GRBs expected to be detected by the IPN in this area during the 2\,d window is 0.09, so the association is fairly secure.
Unfortunately no redshift measurement was obtained of this event.

\subsubsection{ZTF\,21aayokph / AT\,2021lfa}

\begin{figure}[hbt!]
    \centering
    \includegraphics[width=1.0\columnwidth]{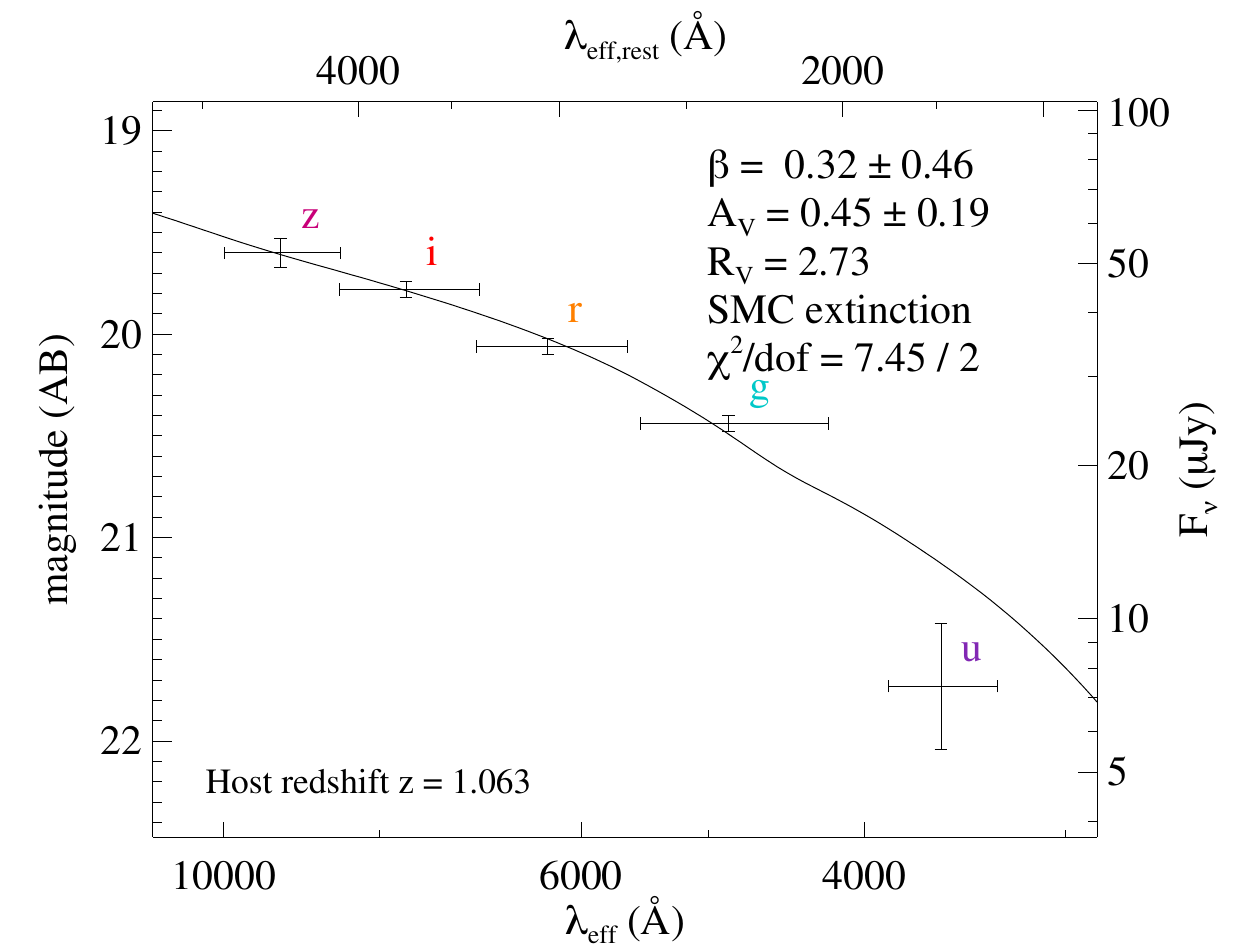}
    \caption{The UVOIR SED of AT\,2021lfa, an afterglow with no associated detected GRB. The SED was obtained at $\Delta t=1.7\,$d by LT in the $ugriz$ bands and shows a dropoff before the $u$ band, likely due to extinction. We show a fit to an SMC dust extinction law.}
    \label{fig:at2021lfa-sed}
\end{figure}

ZTF\,21aayokph was first detected in a public ZTF image on 2021 May 04.23 at $r=18.60\pm0.08\,$mag. The last nondetection was 1.92\,d prior at $r>20.23\,$mag (also in a public-survey image), giving a rise rate in $r$ band of $>0.8\,$mag\,\pday.
The source was detected in both $r$ band and $g$ band the first night, with an extinction-corrected red color of $g-r=0.17\pm0.14\,$mag.
The nearest object in Legacy Survey pre-imaging was 2\farcs9 away, at $g=24.2$\,mag and $r=23.5$\,mag.
Owing to the fast rise, red color, and lack of detected host galaxy at the transient position, the transient was flagged by the daily scanner on May 04.7.
We searched for an associated GRB; none was identified in the 1.92\,d window consistent with the optical-transient position.
We publicly reported the source \citep{Yao2021_ZTF21aayokph} and saved it to the TNS, where it was assigned the name AT\,2021lfa. 

We obtained a long-slit spectrum of AT\,2021lfa with GMOS-S\footnote{Program GS-2021A-Q-124, PI A. Ho} starting on 2021 May 05.18, or $\Delta t=0.95$\,d after the first optical detection.
The observation was performed in the Nod-and-Shuffle mode with a 1\farcs0-wide slit.
We obtained $2\times 450$\,s exposures with each of the B600 and R400 gratings, providing coverage over the range 3620--9540\,\AA. No flux calibration was performed. The spectrum was reduced using the IRAF package for GMOS.
We clearly identified absorption lines of \ion{Mg}{2}$\,\lambda\lambda$2796, 2803 at $z = 1.0624$ in the B600 and R400 spectra. Absorption lines from \ion{Mg}{1} and \ion{Fe}{2} were marginally detected in the spectra at a consistent redshift. \edit1{Using the same approach that we used to measure the redshift of AT\,2021any, we find $z=1.0624 \pm 0.0003$.
In practice this is a lower limit, and} the lack of observed damped Lyman-$\alpha$ places an upper limit of $z<2.3$.
We reported the redshift measurement in \citet{Yao2021_ZTF21aayokph}.
The spectrum with our line identifications is shown in Figure~\ref{fig:spec}, and the measured line strengths are reported in Table~\ref{tab:EW}.

We observed the position of AT\,2021lfa with \emph{Swift}-XRT\footnote{Target ID 14306, PI A. Ho} beginning on May 05.23, $\Delta t=1\,$d after the first optical detection \citep{Ho2021_ZTF21aayokph_XRT}.
We obtained two epochs of 5\,ks exposures separated by 2.5\,d and reduced the data using the online tool developed by the \emph{Swift} team. In the first epoch X-ray emission was detected at the transient position with a count rate of $(9.3\pm0.17)\times10^{-3}\,\psec$, and there was no detection in the second epoch with a 3$\sigma$ upper limit of $<4.4\times10^{-3}\,\psec$.
The observation log is provided in Table~\ref{tab:xray-phot}.
Taking a neutral hydrogen column density of $n_H=2.16\times10^{20}\,\pcmsq$ and assuming a power-law spectrum with photon index $\Gamma=2$, \texttt{webpimms} gives an unabsorbed 0.3--10\,keV flux density of $3.5\times10^{-13}$\,\erg\,\pcmsq\,\psec.
The X-ray measurements are shown in Figure~\ref{fig:orphan-lcs}.

We obtained follow-up observations with the SEDM and LT,
\edit1{which showed rapid fading of 1.9\,mag\,day$^{-1}$ in $r$ band.}
LT image reduction was provided by the basic IO:O pipeline.
P60 and LT image subtraction was performed following \citet{Fremling2016} using PS1 images for $griz$ and SDSS for $u$.
Our photometric observations are provided in Table~5.
Optical photometry was also obtained by other groups \citep{Watson2021_ZTF21aayokph,Butler2021_ZTF21aayokph,Kim2021_ZTF21aayokph,Fu2021_ZTF21aayokph,Oconnor2021_ZTF21aayokph,Pankov2021_ZTF21aayokph,Rossi2021_ZTF21aayokph}, including a claimed detection 3\,hr prior to the first ZTF detection \citep{Lipunov2021_ZTF21aayokph}.
The $r$-band light curve is shown in Figure~\ref{fig:compare-opt-lc-kann}, the multiband optical light curve is shown in Figure~\ref{fig:orphan-lcs},
and the optical SED obtained with LT at $\Delta t=1.7\,$d is shown in Figure~\ref{fig:at2021lfa-sed}.

We obtained nine separate epochs of observations of AT\,2021lfa with the VLA under our ToO program\footnote{VLA/21A-319, PI D. Perley}.
The epochs generally involved various combinations of X-, Ku-, and C-bands, although in one epoch L-band was also included, and in some cases an observation was repeated in the lowest-frequency band after a few hours or days to look for evidence of short-timescale scintillation.
The observation log is provided in Table~\ref{tab:radio-phot}.
We employed 3C286 as a flux calibrator for all observations and J1224+0330 as the phase calibrator.  In three of these epochs we also observed J1407+2827 in X-band only as a polarization leakage calibrator, although no evidence of polarization in the afterglow was detected.  Observations were reduced using \texttt{AIPS} in the same manner as for the observations of AT\,2021any.  
The 10\,GHz radio light curve is shown in Figure~\ref{fig:orphan-lcs},
and the evolution of the SED is shown in Figure~\ref{fig:radio-sed}.

\edit1{We obtained a deep image of the position of AT\,2021lfa using Keck-I/LRIS;
the photometry is reported in Table~\ref{tab:host}.
No host counterpart was detected.}

\subsubsection{ZTF\,21abfmpwn / AT\,2021qbd / GRB\,210610B}

ZTF\,21abfmpwn was first detected in an image obtained as part of the high-cadence partnership survey on 2021 June 11.23 at $g=18.49\pm0.10\,$mag. The last nondetection was 0.85\,d prior at $r>20.20\,$mag (in a public survey image), and the most recent nondetection in the same filter was 0.94\,d prior at $g>20.46\,$mag (also in a public survey image).
The rise time was therefore $>2.1\,$mag\,\pday\ in $g$ band.
The daily scanner promptly identified ZTF\,21abfmpwn as the afterglow to the \emph{Swift} LGRB\,210610B \citep{Page2021_ZTF21abfmpwn}, which was known to have a bright associated X-ray and optical afterglow at $z=1.1345$ \citep{Fynbo2021_ZTF21abfmpwn,deUgartePostigo2021_ZTF21abfmpwn,Dutta2021_ZTF21abfmpwn}.
GRB\,210610B was also detected by \emph{Fermi}-GBM \citep{Malacaria2021} and Konus-\emph{Wind} \citep{Frederiks2021_ZTF21abfmpwn}.
The burst time was 9.7\,hr before the first ZTF detection.

The transient would have passed our selection criteria regardless of the known GRB association owing to its rapid rise, its red color (extinction-corrected $g-r=0.32\pm0.07\,$mag), 
significant intranight fading (0.32\,mag in 3.1\,hr, \edit1{or 2.5\,mag\,\pday}),
and faint ($g=23$, $r=23$\,mag) object 0\farcs3 away in Legacy Survey \edit1{pre-}imaging.
We publicly reported the ZTF detection \citep{Perley2021_ZTF21abfmpwn} and saved the afterglow to the TNS, where it was assigned the name AT\,2021qbd.
We provide the ZTF photometry in Table~\ref{tab:opt-phot}.

Because AT\,2021qbd was associated with a well-observed GRB, we did not obtain any further follow-up observations.
The afterglow was also detected at millimeter \citep{Laskar2021_ZTF21abfmpwn,Smith2021_ZTF21abfmpwn} and radio \citep{Alexander2021_ZTF21abfmpwn} wavelengths.

\section{Comparison to the LGRB Population}
\label{sec:comparison}

In this section we compare the properties of the optically discovered afterglows to the population of optical afterglows detected in follow-up observations of GRB triggers.

\subsection{Redshift Distribution}
\begin{figure}[htb!]
    \centering
    \includegraphics[width=0.9\columnwidth]{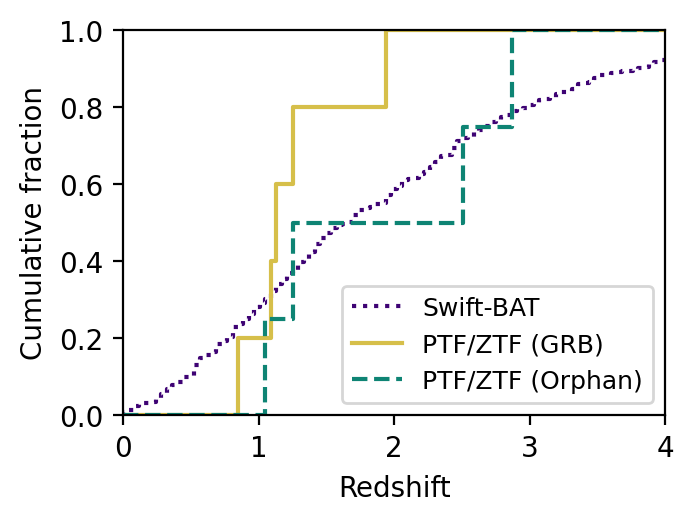}
    \caption{Cumulative redshift distribution of optically discovered afterglows with (five events; solid line) or without (four events; dashed line) associated detected GRBs. For reference we show the distribution for afterglows discovered in follow-up observations to triggers from \emph{Swift}-BAT (dotted line).
    The optically discovered events span a wide range of redshifts ($z=0.9$ to $z=3$) but the small sample size precludes the discernment of any difference in the populations.
    }
    \label{fig:zdist}
\end{figure}

In Figure~\ref{fig:zdist} we show the cumulative redshift distribution of all optically discovered events to date compared to that of
\emph{Swift}-BAT GRBs with optical afterglows\footnote{\url{https://swift.gsfc.nasa.gov/archive/grb\_table/}}.
Current optical searches sample a wide range of redshifts, from $z=0.9$ to $z=3$.
However, we cannot discern any statistically significant differences between the optically selected and GRB-selected events at this stage owing to the small sample size.
In the future, it will be interesting to see whether the ``orphan'' events lie at different redshifts from the events with associated detected GRBs.

\subsection{Prompt Emission}
\label{sec:comparison-grb}

To compare the prompt emission properties of the optically discovered afterglows to the GRB-discovered population,
we begin by calculating the basic properties of the accompanying GRB for each event,
which are summarized in Table~\ref{tab:grb}.
For events with probable associated GRBs (as established in \S\ref{sec:observations}),
we calculate the time interval containing 5\% to 95\% of the burst fluence ($T_{90}$), the fluence, the isotropic-equivalent $\gamma$-ray energy release $E_{\gamma,\mathrm{iso}}$ and luminosity $L_{\gamma,\mathrm{iso}}$, and the peak energy $E_p$.
Because all the bursts were detected by Konus-\emph{Wind},
we use the same approach as that used by \citet{Tsvetkova2017} and \citet{Tsvetkova2021}.
For bursts also detected by \emph{Fermi}-GBM, we confirmed that the energetics inferred from \emph{Fermi}-GBM data are consistent with the Konus-\emph{Wind} values.
We recalculated these values \edit1{even} for \edit1{previously} published events in order to provide consistent measurements.

\begin{deluxetable*}{lrrrrrrrr} 
\tablecaption{Properties of (or limits on) the prompt $\gamma$-ray emission accompanying afterglows discovered by optical surveys$^{a}$.  \label{tab:grb}} 
\tablewidth{0pt} 
\tablehead{ \colhead{Name (GRB)} & \colhead{Inst.$^{b}$} & \colhead{$T_{90}^c$} & \colhead{Fluence} & \colhead{Peak Flux} & \colhead{$E_{p}^{c,d}$} & \colhead{$E_{\gamma,\mathrm{iso}}^e$} & \colhead{$L_{\gamma,\mathrm{iso}}^e$} & \colhead{Ref.$^f$}\\[-0.3cm] \colhead{} & \colhead{} & \colhead{(s)} & \colhead{($10^{-6}\,\erg\,\pcmsq$)} & \colhead{($10^{-6}\,\erg\,\pcmsq\,\psec$)} & \colhead{(keV)} & \colhead{($10^{52}$\,erg)} & \colhead{($10^{52}$\,erg\,\psec)} & \colhead{}} 
\tabletypesize{\scriptsize} 
\startdata 
PTF11agg  & -- & -- & $<0.4$ & $<0.4$ & -- & -- & -- & \\
iPTF14yb (140226A) & IKO & $13.5 \pm 1.1$$^{g}$ & $5.1^{+5.6}_{-0.9}$ & $0.64^{+0.72}_{-0.14}$ & $453^{+1010}_{-198}$ & $5.4^{+6.0}_{-1.0}$ & $2.0^{+2.3}_{-0.4}$ & \\
ATLAS17aeu (170105A) & AIKP & $3.3 \pm 0.6$$^{h}$ & $2.39^{+0.12}_{-0.15}$ & $0.75^{+0.04}_{-0.05}$ & $56^{+1}_{-12}$ & -- & -- & \\
AT2019pim$^i$  & -- & -- & $<0.35$ & $<0.04$ & -- & $<0.3$ & $<0.03$ & [1]\\
AT2020blt  & -- & -- & $<0.4$ & $<0.2$ & -- & $<1.0$ & $<1.3$ & this work; [2]\\
AT2020kym (200524A) & AFK & $66.0 \pm 9.9$ & $34.8^{+4.3}_{-3.9}$ & $6.33^{+1.04}_{-0.99}$ & $215^{+28}_{-27}$ & $13.6^{+1.7}_{-1.5}$ & $5.6^{+0.9}_{-0.9}$ & \\
AT2020sev (200817A) & FIKS & $388.6 \pm 26.5$ & $6.6^{+4.3}_{-0.9}$ & $0.38^{+0.26}_{-0.07}$ & $427^{+812}_{-171}$ & -- & -- & \\
AT2020yxz (201103B) & GIK & $63.6 \pm 20.2$$^{j}$ & $52.6^{+5.0}_{-4.7}$ & $14.9^{+1.4}_{-1.4}$ & $403^{+44}_{-39}$ & $16.9^{+1.6}_{-1.5}$ & $10.1^{+0.9}_{-0.9}$ & \\
AT2021any  & -- & -- & $<10.0$ & $<1.0$ & -- & $<14.3$ & $<5.0$ & \\
AT2021buv (210204A) & ABCFIKOS & $197.0 \pm 6.0$ & $80.9^{+5.4}_{-6.8}$ & $5.6^{+1.06}_{-1.05}$ & $137^{+13}_{-11}$ & $22.6^{+1.5}_{-1.9}$ & $2.9^{+0.6}_{-0.5}$ & \\
AT2021cwd (210212B) & IK & $41.2 \pm 2.9$$^{j}$ & $8.7^{+1.7}_{-1.1}$ & $0.67^{+0.48}_{-0.16}$ & $208^{+84}_{-45}$ & -- & -- & \\
AT2021lfa  & -- & -- & $<0.4$ & $<0.4$ & -- & $<0.12$ & $<0.26$ & \\
AT2021qbd (210610B) & FKS & $48.5 \pm 4.2$ & $136.0^{+5.5}_{-5.4}$ & $10.7^{+1.11}_{-1.07}$ & $255^{+8}_{-8}$ & $47.8^{+1.9}_{-1.9}$ & $8.0^{+0.8}_{-0.8}$ & \\
\enddata 
\tablenotetext{a}{Uncertainties are given at 1$\sigma$ confidence. Values were calculated in the 20\,keV--10\,MeV range             (80--1500\,keV for $T_{90}$) unless otherwise specified.}
\tablenotetext{b}{A: \emph{ASTROSAT} (CZTI); B: \emph{GECAM-B} (GRD);  C: \emph{CALET} (GBM); F: \emph{Fermi} (GBM); G: \emph{AGILE} (MCAL); I: \emph{INTEGRAL} (SPI-ACS); K: \emph{Konus} (Wind); O: \emph{Mars-Odyssey} (HEND); P: \emph{TG2} (POLAR); S: \emph{Swift} (BAT).}
\tablenotetext{c}{$T_{90}$ and $E_p$ values are presented in the observer frame.}
\tablenotetext{d}{$E_p$ was measured using the time-integrated spectrum.}
\tablenotetext{e}{$E_{\gamma,\mathrm{iso}}$ and $L_{\gamma,\mathrm{iso}}$ have             a $K$-correction applied, which transforms the energetics             from the observer-frame 20\,keV--10\,MeV energy range to the             $1/(1+z)$\,keV--$10/(1+z)$\,MeV band.}
\tablenotetext{f}{Quantities were calculated as part of this work unless specified in the Ref. column.}
\tablenotetext{g}{For GRB\,140226A the $T_{90}$ was calculated using \emph{INTEGRAL} (SPI-ACS) data at $\gtrsim80\,$keV.}
\tablenotetext{h}{The $T_{90}$ for GRB\,170105A is calculated in the 70--300\,keV Konus-\emph{Wind} energy band and is consistent with \citet{Stalder2017}. In the softer 20--70\,keV band, $T_{90}=20\pm4$\,s (more similar to \citealt{Bhalerao2017}).}
\tablenotetext{i}{The upper limits are measured from a \emph{Fermi}-GBM targeted search, using a soft template with 10\,s duration.}
\tablenotetext{j}{For GRB\,201103B and GRB\,210212B, T90 is calculated in the 80--1000\,keV band.}
\tablerefs{[1] Perley et al. (in prep.), [2] \citet{Ho2020d}.} 
\end{deluxetable*}

For orphan events, we estimate the time of an associated burst using a power-law fit to the optical light curve, and use the coverage by high-energy satellites during the relevant period to set limits on the properties of an associated GRB.
Because \emph{Fermi}-GBM and Konus-\emph{Wind} operate using a flux-based trigger,
we use the estimated flux sensitivity to set an upper limit on $L_{\gamma,\mathrm{iso}}$.
However, because $E_{\gamma,\mathrm{iso}}$ is an important physical quantity for drawing comparisons with the LGRB population, 
we also use a typical fluence threshold to estimate an upper limit on $E_{\gamma,\mathrm{iso}}$.
The upper limits on $L_{\gamma,\mathrm{iso}}$ and $E_{\gamma,\mathrm{iso}}$ in Table~\ref{tab:grb} have cosmological corrections applied.

We fit the AT\,2021any ZTF light curve with a broken power law following the procedure applied to AT\,2020blt \citep{Ho2020d}, which made use of a modified fitting function from \citet{Zeh2006}.
We fit the function to the $g$-band and $r$-band light curves simultaneously, 
with a constant offset in magnitude between the light curves.
Since the host galaxy is not detected in deep Legacy Survey \edit1{pre-}imaging,
and the ZTF points are all brighter than 20\,mag,
we do not fit for a constant offset from the host.
Using the $\texttt{curve\_fit}$ package in $\texttt{scipy}$,
we find a best-fit $t_0$ that is 9\,min after the last nondetection and 11\,min prior to the first detection.

The best-fit temporal power-law index is $\alpha_1=0.57\pm0.04$ prior to the break,
and $\alpha_2=1.18\pm0.01$ after the break, with a break at $\Delta t=3.9$\,hr.
The goodness of fit is $\chi^2/\nu=3.4$ for $\nu=8$ degrees of freedom.
The pre-break index we measure is shallower than the value of $\alpha_1=0.89\pm0.03$ reported by \citet{Gupta2021} on the basis of GCN photometry.
The post-break index we measure is shallower than the value of $\alpha_2=2.30$ reported by \citet{Kann2021_ZTF21aaeyldq_jet_break}.
The best fit is shown in Figure~\ref{fig:orphan-lcs}.

We fit the AT\,2021lfa light curve to a single power law, to the $g$, $r$, and $i$ light curves simultaneously, 
with a constant offset in magnitude between each pair of bands.
Since the host galaxy is not detected in deep Legacy Survey \edit1{pre-}imaging,
we do not fit for a constant offset from the host.
We find a best-fit $t_0$ of May 03.29, which is 0.98\,d prior to the first detection and 0.94\,d after the last nondetection.
The best-fit temporal power-law index is $\alpha=2.54\pm0.02$, with a goodness of fit of $\chi^2/\nu=4.7$ for $\nu=18$ degrees of freedom.
The fits do not capture an apparent late-time flattening in the $r$-band and $i$-band light curves.
The fit is shown in Figure~\ref{fig:orphan-lcs}.

\begin{figure*}[!ht]
    \centering
    \includegraphics[width=0.6\textwidth]{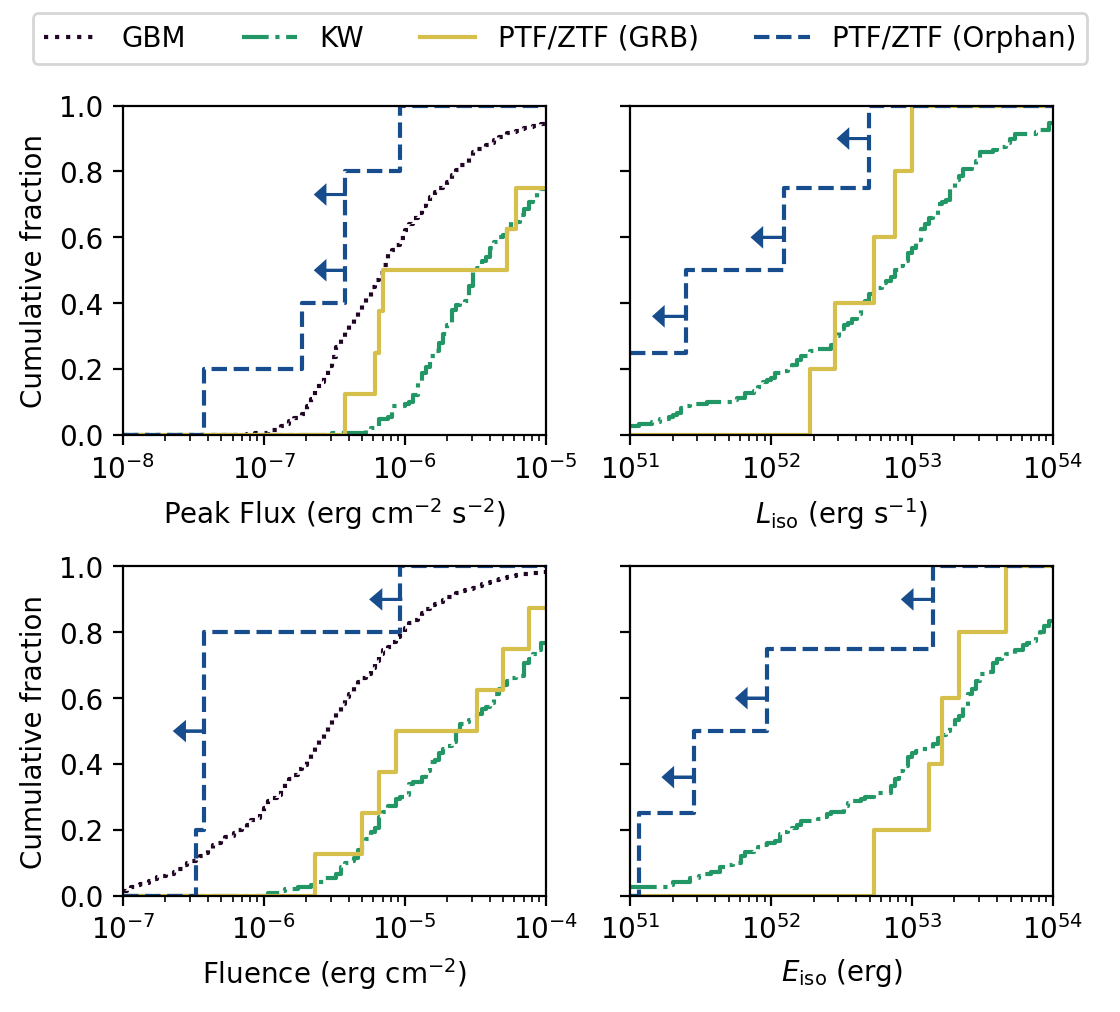}
    \caption{Cumulative flux (top left), $L_\mathrm{iso}$ (top right), fluence (bottom left), and $E_\mathrm{iso}$ (bottom right) distributions of optically discovered afterglows: events with detected GRBs (solid line) and orphan events (dashed line). For the orphan events, we set the values equal to the upper limit.
    For comparison, we show the distributions of GRB-selected events from GBM (dotted) and Konus-\emph{Wind} (dash-dot).
    The sample size is small, but it appears that LGRBs accompanying the orphan events would have to be less luminous and less energetic than the detected LGRB counterparts. 
    }
    \label{fig:fluence-dist}
\end{figure*}

Our best-fit $t_0$ for AT\,2021any is within the interval of visibility to \emph{Fermi}-GBM. Using the \emph{Fermi}-GBM trigger sensitivity, we estimate that the upper limit on the peak flux is $1\times10^{-7}\,\erg\,\psec\,\pcmsq$, although
a ground-based analysis could reduce the sensitivity by a factor of 2--3.
However, early optical afterglow light curves can have complex behavior \citep{Kann2010}, so our power-law fit may not be appropriate.
To be conservative, we report an upper limit based on Konus-\emph{Wind}.
AT\,2021any had only nonstandard Konus-\emph{Wind} data, so we estimate an upper limit on the flux of $10^{-6}\,\erg\,\pcmsq\,\psec$ (20--10,000\,keV; 3.68\,s timescale),
a factor of a few higher than typical Konus-\emph{Wind} peak flux upper limits \citep{Ridnaia2020_GW}.
For the fluence, a conservative upper limit is $10^{-5}\,\erg\,\pcmsq$ \citep{Tsvetkova2021}.
The corresponding limits on the energetics, with cosmological corrections, is $L_\mathrm{iso} < 5.0\times10^{52}\,\erg\,\psec$ and $E_\mathrm{iso} < 14.3 \times 10^{52}\,\erg$.

For AT\,2021lfa, the steep power-law index of the optical light curve suggests that we are observing after the jet break (as discussed in \S\ref{sec:orphans}),
so we are unable to use our estimated $t_0$ as a burst time.
Given the large window between the last nondetection and the first detection,
we use the upper limit on the peak flux from Konus-\emph{Wind} data of $4\times10^{-7}\,\erg\,\pcmsq\,\psec$ (20--10,000\,keV; 2.944\,s timescale),
giving $L_{\gamma,\mathrm{iso}}<2.6\times10^{51}\,\erg\,\psec$.
To set an upper limit on the fluence, we searched for a GRB counterpart using the procedure of \citet{Tsvetkova2021}.
The nondetection suggests that the limiting fluence is comparable to the fluence of the weakest burst from \citet{Tsvetkova2021}, so approximately $4\times10^{-7}\,\erg\,\pcmsq$ (20\,keV--10\,MeV).
The corresponding upper limits on $E_\mathrm{iso}$ and $L_\mathrm{iso}$ are reported in Table~\ref{tab:grb}.

We compute new upper limits for PTF\,11agg in a consistent way. The interval of interest is the 21.5\,hr between the last nondetection and the first detection.
Konus-\emph{Wind} was taking data with stable background conditions for 98\% of the time.
We find an upper limit on the peak flux of $4\times10^{-7}\,\erg\,\pcmsq\,\psec$ (20\,keV--10\,MeV, 2.944\,s timescale), which is insensitive to the assumed spectral model.
We set a fluence threshold in the same way as for AT\,2021lfa.

\begin{deluxetable*}{lrrrrrrrr}[htb!]
\tablecaption{Properties of the optical light curves for optically discovered afterglows with redshift measurements.
\label{tab:opt-lc}} 
\tablewidth{0pt} 
\tablehead{ \colhead{Name} & \colhead{$\alpha_O^a$} & \colhead{$L_u^{b}$ (11h)} & \colhead{$L_u$ (1d)} & \colhead{Ref.$^c$} \\[-0.3cm]
\colhead{} & \colhead{} & \colhead{(erg\,\psec)} & \colhead{(erg\,\psec)} & \colhead{}
} 
\tabletypesize{\scriptsize} 
\startdata 
iPTF14yb & 1.02 & $8.2\times10^{43}$ & $3.4\times10^{43}$ & [1] \\
AT\,2019pim & 0.9 ($\Delta t^{d}<3\,$d), 0 ($3\,\days<\Delta t<8\,$d), 2 ($\Delta t>8\,\days$) & $4.4\times10^{44}$ & $1.6\times10^{44}$ & [2] \\
AT\,2020blt & 0.54 ($\Delta t<1\,$d), 2.62 ($\Delta t>1\,$d) & $8.4\times10^{44}$ & $1.9\times10^{44}$ & [3] \\
AT\,2020kym & 1.53 ($\Delta t<0.8\,$d), 0.8 ($\Delta t>0.8\,$d) & $2.5\times10^{44}$ & $1.0\times10^{44}$ & \\
AT\,2020yxz & 0.96 & $1.2\times10^{45}$ & $4.2\times10^{44}$ & [4] \\
AT\,2021any & 0.7 ($\Delta t<0.8\,$d), 2.3 ($\Delta t>0.8\,$d) & $4.0\times10^{44}$ & $1.9\times10^{44}$ & [5] \\
AT\,2021buv & 0.6 ($\Delta t<2.2\,$d), 1.7 ($\Delta t>2.2\,$d) & $1.9\times10^{45}$ & $4.3\times10^{44}$ & [6] \\
AT\,2021lfa & 2.54 & $2.4\times10^{45}$ & $3.9\times10^{44}$ & \\ 
AT\,2021qbd & 1.57 & $2.0\times10^{45}$ & $3.3\times10^{44}$ & [7] \\
\enddata 
\tablerefs{
[1] \citet{Cenko2015},
[2] Perley et al. in prep.
[3] \citet{Ho2020d}
[4] \citet{Andreoni2021}
[5] \citet{Kann2021_ZTF21aaeyldq_jet_break}
[6] \citet{Rossi2021}
[7] \citet{Pankov2021_210610B}
}
\tablenotetext{a}{Optical temporal power-law decay index.}
\tablenotetext{b}{Rest-frame $u$-band luminosity.}
\tablenotetext{c}{Reference for the temporal power-law index.}
\tablenotetext{d}{Time ranges are in the observer frame.}
\end{deluxetable*} 

We can now compare the energetics of the prompt high-energy emission to that of typical LGRBs.
The burst durations and $L_\mathrm{iso}$ values in Table~\ref{tab:grb} classify each event as a classical LGRB with $L_{\mathrm{iso}}>10^{49.5}\,\erg\,\psec$ \citep{Cano2017}. The values of $T_{90}$ range from 3.3\,s (for ATLAS17aeu / GRB\,170105A) to 388.6\,s (for AT\,2020sev / GRB\,200817A).
It is not surprising that we would detect more LGRB afterglows than short-duration GRB (SGRB) afterglows,
because LGRB afterglows are an order of magnitude more luminous than SGRB afterglows \citep{Kann2011,Berger2014}.

For the orphan events we cannot rule out an associated prompt LGRB with a ``classical'' high luminosity of $L_{\gamma,\mathrm{iso}} > 10^{49.5}\,\erg\,\psec$ \citep{Cano2017}.
This can be understood as follows. For most events, the coverage by sensitive detectors such as \emph{Swift}-BAT and \emph{Fermi}-GBM, together with the uncertainty on the burst time, means that the most robust upper limit comes from Konus-\emph{Wind}.
The Konus-\emph{Wind} flux threshold of a few
$\times10^{-7}\,\erg\,\pcmsq\,\psec$ could only rule out a $10^{49.5}\,\erg\,\psec$ GRB at $z=0.16$ (800\,Mpc), a much lower redshift than any of our events.

The cumulative distributions of the prompt energetics are shown in Figure~\ref{fig:fluence-dist}.
For comparison, we display the distribution of GRBs in the \emph{Fermi}-GBM Burst Catalog\footnote{\url{https://heasarc.gsfc.nasa.gov/W3Browse/fermi/fermigbrst.html}} \citep{Gruber2014,vonKienlin2014,NarayanaBhat2016,vonKienlin2020,Poolakkil2021},
and LGRBs from Konus-\emph{Wind} with redshift measurements \citep{Tsvetkova2017}.
The sample size is too small to discern statistically significant differences.
However, so far it appears that if the orphan events are ordinary LGRBs, the bursts have lower luminosities and energies than the events with detected LGRBs.
The limits are consistent with a picture in which the GRB-associated events have higher luminosities and energies, making them detectable by all-sky GRB monitors, whereas the orphan events have lower but still ordinary LGRB parameters.

\subsection{Optical Light Curves}
\label{sec:comparison-opt-lc}

In Figure~\ref{fig:compare-opt-lc-kann} we show the $r$-band light curves of the ZTF afterglows compared to the sample of optical afterglows detected in follow-up observations of \emph{Swift}-BAT GRBs presented by \citet{Kann2010},
and of \emph{Fermi}-GBM GRBs presented by \citet{Singer2015}.
The shaded region indicates the 25th and 75th percentile bounds of the \citet{Kann2010} sample.
The left panel shows the events with GRB associations, including two events detected prior to ZTF (iPTF14yb and ATLAS17aeu), and the right panel shows the events with no associated detected GRB, including one event detected prior to ZTF (PTF11agg).
The left panel of Figure~\ref{fig:compare-opt-lc-kann} demonstrates that the events with detected GRBs have particularly bright afterglows --- most events are brighter than the 75th percentile from the \citet{Kann2010} sample.

Comparing the brightness of the orphan optical afterglows, shown in the right panel of Figure~\ref{fig:compare-opt-lc-kann}, is challenging because of the uncertainty in the true time of first light.
Using the best-fit $t_0$ values for PTF11agg and AT\,2021lfa, the light curves appear to be brighter than those of most GRB afterglows.
The best-fit $t_0$ values for AT\,2021any and AT\,2020blt suggest that the light curves are fairly typical in brightness for GRB afterglows.
The most extreme case --- a value of $t_0$ equal to the last nondetection --- would put the first AT\,2020blt detection at $\Delta t=0.74$\,d,
and the first AT\,2021any detection at $\Delta t=0.014$\,d.
So, while AT\,2020blt could also have been fairly bright, AT\,2021any would still be typical.

\begin{figure}
    \centering
    \includegraphics[width=0.8\columnwidth]{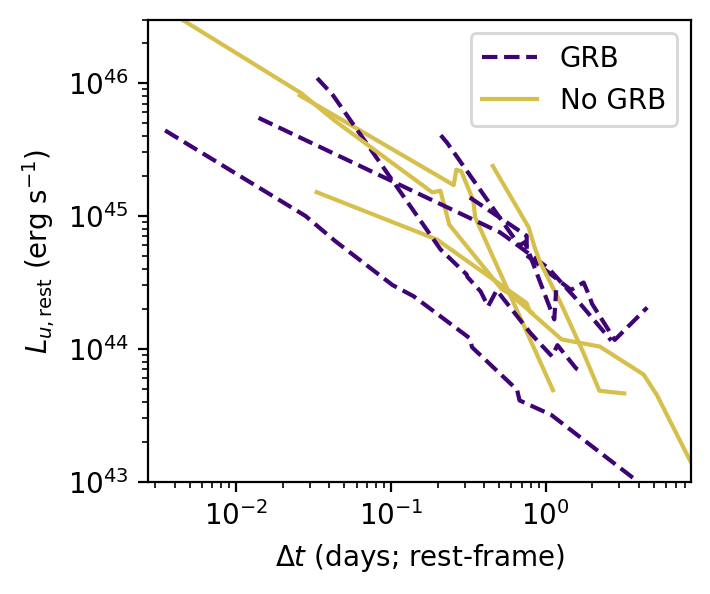}
    \caption{Rest-frame $u$-band light curves of all optically discovered afterglows with redshift measurements, including events with (dashed line) and without (solid line) associated detected GRBs. There is no clear difference between the GRB-associated and orphan events.}
    \label{fig:lc-rest}
\end{figure}

For events with measured redshifts, we construct rest-frame $u$-band optical light curves to compare the optical luminosity of LGRB afterglows.
We convert the observed $r$-band light curves using

\begin{equation}
\label{eq:opt-lc}
    L_u (t) = 4 \pi D_L^2 F_r(t) (1+z)^{-\alpha_{\rm O}+\beta_{\rm O}-1} \left( \frac{\nu_u}{\nu_r} \right)^{-\beta_{\rm O}},
\end{equation}

\noindent where $D_L$ is the luminosity distance, $F_r(t)$ is the observed $r$-band flux at a given time $t$, and $z$ is the redshift. The temporal and spectral indices are defined as $F_\nu \propto t^{-\alpha_{\rm O}}\nu^{-\beta_{\rm O}}$. The frequencies are set to $\nu_u=8.3\times10^{14}\,$Hz and $\nu_r=4.9\times10^{14}\,$Hz for $u$ band and $r$, respectively.
For each event we adopt a typical value of $\beta_{\rm O}=0.6$ \citep{Greiner2011}. The values of $\alpha_{\rm O}$ we adopt for each burst are shown in Table~\ref{tab:opt-lc}, and the resulting light curves are shown in Figure~\ref{fig:lc-rest}.
We also take a cross-section of the light curves at 11\,hr and 1\,d in the rest frame; we provide the luminosities in Table~\ref{tab:opt-lc}.
The luminosities are typical of LGRB light curves \citep{Racusin2011} and we find no evidence that the orphan events are distinct.

\begin{figure}
    \centering
    \includegraphics[width=0.7\columnwidth]{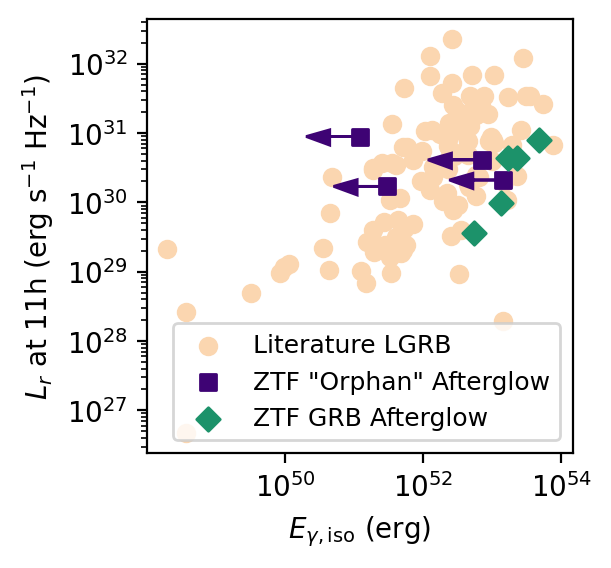}
    \caption{Rest-frame $r$-band luminosities at 11\,hr for all optically discovered afterglows with redshift measurements, compared to the LGRB sample from \citet{Nysewander2009}.
    We cannot rule out an associated LGRB for any of the orphan events.}
    \label{fig:nysewander}
\end{figure}

As a final check, we calculate the rest-frame $r$-band luminosities of the orphan events at 11\,hr, and plot them along with the limits on $E_\mathrm{iso}$ in Figure~\ref{fig:nysewander}.
We compare them to a sample of LGRBs from \citet{Nysewander2009}.
For the comparison, we calculate the rest-frame $r$-band luminosity using the same adopted spectral index ($\beta=1$) as in \citet{Nysewander2009}.
The limits on $E_\mathrm{iso}$ are not sensitive enough to rule out a contemporaneous LGRB for any event.

\subsection{Multiwavelength Properties of the Orphan Afterglows}
\label{sec:orphans}

In this section, we analyze the multiwavelength data of AT\,2021any and AT\,2021lfa, and put the properties of all optical orphan afterglows discovered to date in the context of the LGRB population.
Thus far, five orphan optical afterglows have been discovered:
PTF11agg ($0.5 \lesssim z \lesssim 3.0$), AT\,2019pim ($z=1.2596$), AT\,2020blt ($z=2.9$), AT\,2021any ($z=2.5131$), and AT\,2021lfa ($z=1.0624$).
The properties of PTF11agg and AT\,2020blt were presented by \citet{Cenko2013} and \citet{Ho2020d} respectively, and AT\,2019pim will be discussed in detail in a separate paper by Perley et al.
\citet{Gupta2021} analyzed the public data of AT\,2021any and concluded that its properties were consistent with being a LGRB viewed on-axis.
By modeling the optical light curve of AT\,2020blt, and adopting deeper limits on accompanying $\gamma$-ray emission than did \citet{Ho2020d}, \citet{Sarin2021} argued that it was likely viewed on-axis but with an unusually low $\gamma$-ray efficiency of $<2.8\%$.

The optical light curves of AT\,2021any and AT\,2021lfa have temporal indices that are similar to those of LGRB optical afterglows in the literature \citep{Zeh2006,Kann2010,Li2012,Wang2018}.
The AT\,2021lfa power-law index of $\alpha=2.54\pm0.02$ is close to expectations for synchrotron emission from a power-law distribution of electrons after the edge of the jet is visible, for a typical electron energy power-law index of $p=2.5$ \citep{Sari1999}.
The AT\,2021any light curve shows a clear break.
The post-break index of $\alpha_2=1.2$ is close to expectations for synchrotron emission before the edge of the jet is visible \citep{Sari1999}.
The pre-break index of $\alpha_1=0.6$ is quite shallow, and close to the pre-break value measured for AT\,2020blt \citep{Ho2020d}.

\edit1{Light-curve breaks} are commonly attributed to two effects in collimated relativistic jets \citep{Rhoads1997,Sari1999}: sideways expansion and the edge of the jet being visible, both of which are thought to occur when the Lorentz factor $\Gamma(t) \propto \theta_0^{-1}$, where $\theta_0$ is the opening angle of the jet.
The fact that we did not observe the break in AT\,2021lfa suggests that it occurred within a day of the burst in the rest frame, which is common for LGRB optical afterglows \citep{Zeh2006,Kann2010,Wang2018}, and enables us to estimate the opening angle of the jet.
We use the expression from \citet{Sari1999},

\begin{equation}
\label{eq:jetbreak}
    t_\mathrm{jet} = 6.2 \left( \frac{E_{52}}{n_1} \right)^{1/3} \left( \frac{\theta_0}{0.1} \right)^{8/3}\,\mathrm{hr},
\end{equation}

\noindent where $E_{52}$ is $E_\mathrm{iso}$ in units of $10^{52}\,\erg$, $n_1$ is the circumburst density in units of \pcmcub, and $t_\mathrm{jet}$ is the time of the jet break.
For AT\,2021lfa, we adopt $t_\mathrm{jet}<22\,$hr and $E_{52}<0.12$.
The opening angle, which is more sensitive to the timing of the break, is therefore $<12^{\circ}$.
For AT\,2021any, the shallower decay index suggests that we did not observe the transition.
Taking $t_\mathrm{jet}>0.6\,$d in the rest frame, and $E_{52}<14.3$, we infer an opening angle of $>6^{\circ}$. 

The optical light curves constrain our viewing angle.
For AT\,2021any, the shallow ($\alpha=1.2$) index is consistent with an event viewed directly on-axis, in agreement with \citet{Gupta2021}.
For AT\,2021lfa, from the available photometry we cannot determine whether the event was initially viewed on-axis.
However, given the high luminosity, it must have been viewed at least very close to on-axis.

Interestingly, of the five orphan optical afterglows observed so far, all show either a prominent break in the light curve or \edit1{a steep power-law index consistent with post-break evolution.}
We discussed AT\,2021any and AT\,2021lfa above.
AT\,2020blt had a clear jet break \citep{Ho2020d}.
\edit1{As will be discussed by Perley et al. in prep.,}
AT\,2019pim had a complicated light curve, with several segments having different decay indices: an early segment had $\alpha=0.9$ while a later segment appeared to show a steeper value of $\alpha=2$, which could also represent a break.
The light curve of PTF11agg was fit by a single power law with index $\alpha=1.66\pm0.35$, but possibly a value as steep as $\alpha=2.5$ owing to the uncertainty in the burst time.
The data are consistent with AT\,2021any, AT\,2019pim, and AT\,2020blt being viewed within the initial opening angle of the jet,
while AT\,2021lfa and PTF11agg may have been viewed slightly off-axis.

The optical spectra of AT\,2021any and AT\,2021lfa are also typical of LGRB afterglows.
We compare the strength of the redshift-corrected spectral features of each object with those of a large sample of LGRB afterglow spectra. To do this, we calculate a line-strength parameter (LSP; \citealt{deUgartePostigo2012}) and construct a line-strength diagram. \edit1{The LSP is defined in Equation 1 of \citealt{deUgartePostigo2012}, which we reproduce here:}

\begin{equation}
    \mathrm{LSP} = \frac{1}{N} \sum^{N}_{i=1} \frac{\log EW_i-\langle \log EW\rangle_i}{\sigma_{\mathrm{log}EW,i}},
\end{equation}

\noindent
\edit1{where $EW_i$ is the equivalent width of the spectral lines. Essentially, the LSP quantifies the difference between the strength of absorption features in a given spectrum to the strength of an average GRB spectrum: a positive and negative value means stronger and weaker lines, respectively.
By calculating an LSP and producing a line-strength diagram,
we can identify differences between the composition or ionization of an individual object's environment and the values typical of LGRB environments.}

\begin{figure}
   \centering
   \includegraphics[width=\hsize]{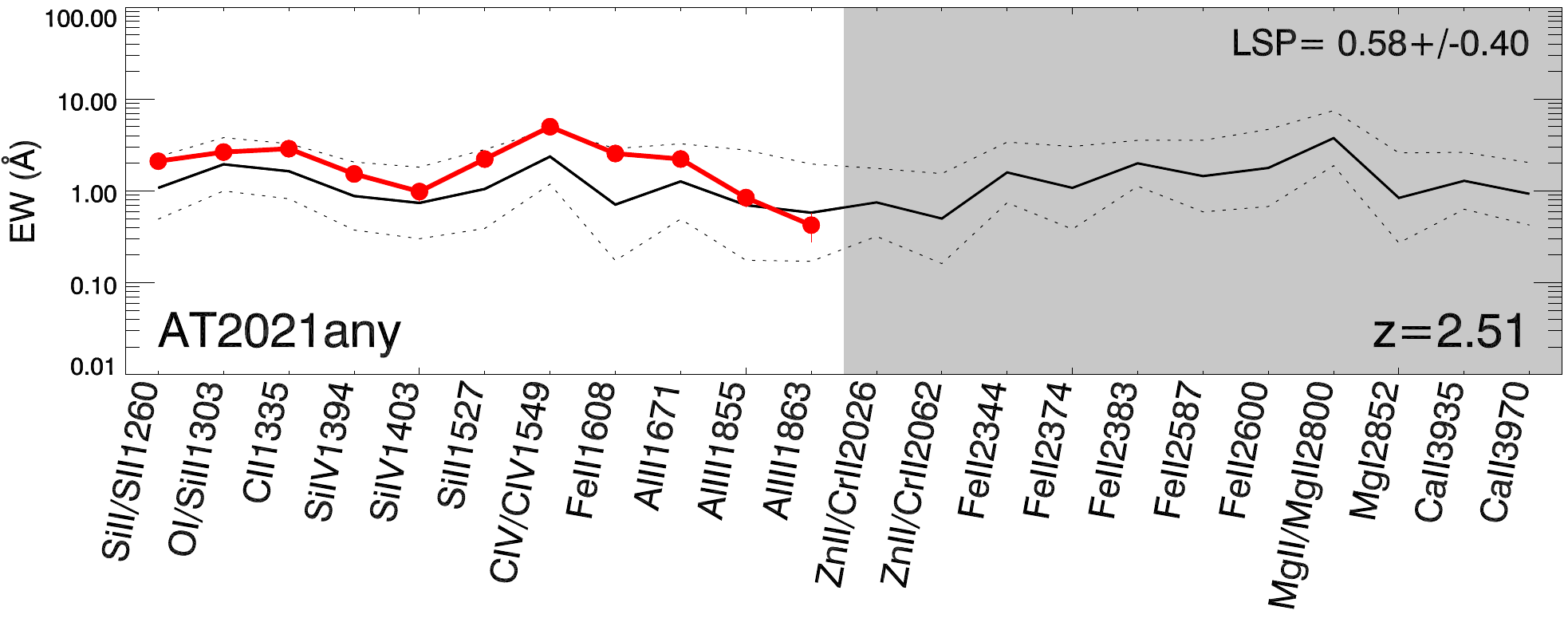}
   \includegraphics[width=\hsize]{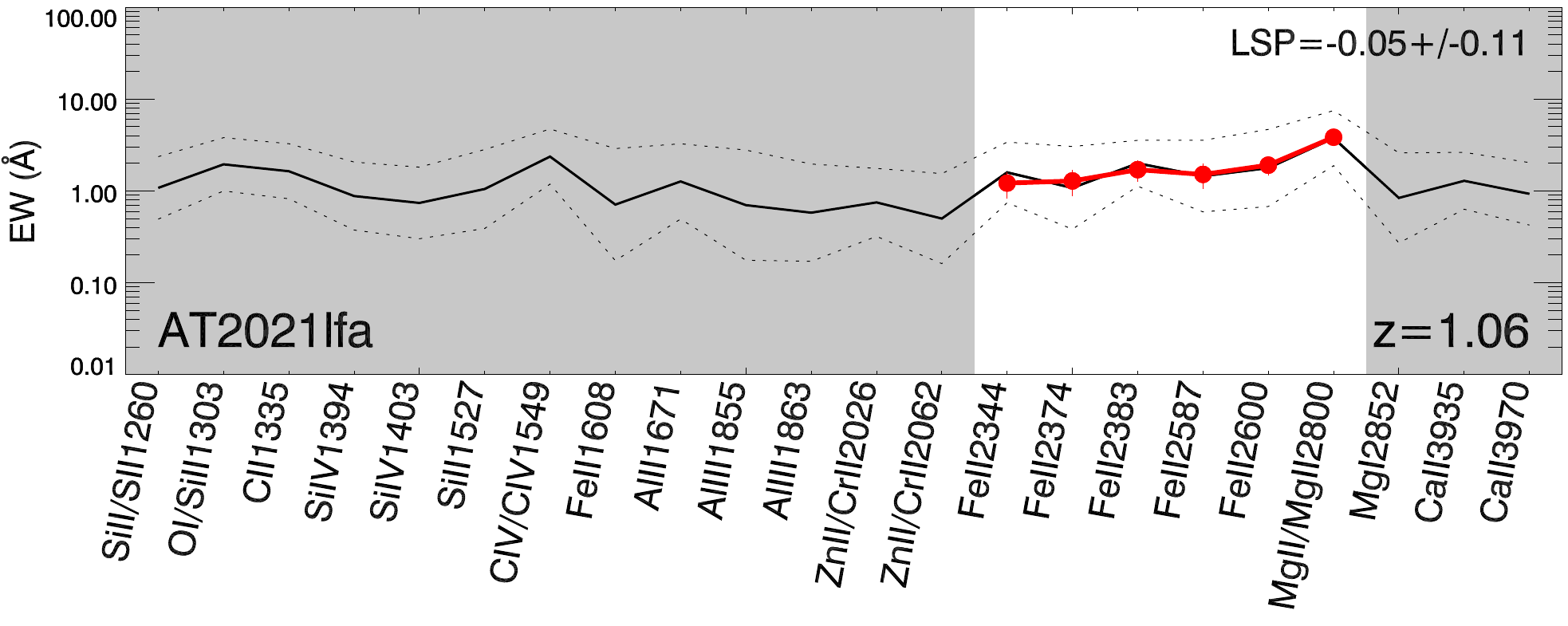}
      \caption{Line-strength diagram comparing the features in the spectra of AT\,2021any and AT\,2021lfa, two apparently orphan afterglows, with those of a large sample of GRB afterglows. The lines of AT\,2021any are stronger than average, but they show similar relative strengths as those of the sample.
      \edit1{The line strengths of AT\,2021lfa are average for LGRB afterglows.}
              }
         \label{fig:LSD}
   \end{figure}

In the case of AT\,2021any, we obtain an LSP of $0.58\pm0.40$, which implies that the observed features are stronger than those of 80\% of the sample. The line-strength diagram for this spectrum (Figure~\ref{fig:LSD}) shows no clear deviations of the relative line strength of the lines with those of the sample, except perhaps a slight deficit of \ion{Al}{3}. So although the lines are stronger than average, there is no significant difference in the ionization state or composition.
For AT\,2021lfa we calculate a value of LSP = $-0.05\pm0.11$, close to the average of the LGRB sample.  
The line-strength diagram (Figure~\ref{fig:LSD}) also shows a line-strength distribution that follows the average values of the sample, with the same relative strength of lines.

The optical to X-ray SEDs of AT\,2021any and AT\,2021lfa are also typical of LGRBs.
For AT\,2021any we take the NOT photometry from \citet{Zhu2021_ZTF21aaeyldq} at $\Delta t=0.8\,$d, and find a spectral index of $\beta_{\rm O}=1.00\pm0.01$ across the $gri$ bands where $f_\nu \propto \nu^{-\beta_{\rm O}}$.
At this epoch, the spectral index from the optical to X-ray band is shallower, $\beta_{\rm OX}=0.6$.
We show the ultraviolet-optical-infrared (UVOIR) SED of AT\,2021lfa in Figure~\ref{fig:at2021lfa-sed}. The best-fit observed spectral index is $\beta_{\rm O}=1.24\pm0.01$, including a rapid drop beteween the $g$ and $u$ bands, which implies host dust extinction.
By fitting an SMC-like extinction law similar to that of the \citep[SMC;][]{Fitzpatrick1999,Gordon2003}, we find $\beta_{\rm O}=0.32\pm0.46$ and $A_V=0.45\pm0.19\,$mag.
At the epoch of the SEDM observation 0.3\,d later, the optical to X-ray index is $\beta_{OX}=0.4$.
The values of $\beta_{\rm O}$, $\beta_{\rm OX}$, and $A_V$ are all standard for LGRB afterglows \citep{Cenko2009,Greiner2011}.

The low signal-to-noise ratio of our X-ray observations precludes a detailed analysis of the X-ray data, so we focus on comparing the overall X-ray luminosity to that of LGRB afterglows.
For comparison we use the analysis of \emph{Swift}/XRT data by \citet{Margutti2013}.
We estimate the rest-frame 0.3--30\,keV luminosity of each of our events, converting the count rate to unabsorbed 0.3--30\,keV flux using \texttt{webpimms}.
We find $L_X=2\times10^{46}\,\erg\,\psec$ for AT\,2021any and $L_X=3\times10^{45}\,\erg\,\psec$ for AT\,2021lfa.
These values are within the typical range for LGRB afterglows at these epochs.
We can set upper limits of $t^{-1.0}$ and $t^{-0.3}$ on the fade rate for AT\,2021any and AT\,2021lfa, respectively, which are consistent with the typical $t^{-1.2}$ in the epochs relevant to our observations \citep{Margutti2013}, and also with the fact that our X-ray data were obtained post-break for AT\,2021lfa.
Since a jet break should be achromatic \citep{Rhoads1999}, the X-ray light curve would be expected to be steep at this stage.

Our VLA radio observations of AT\,2021lfa and AT\,2021any constitute the first multifrequency radio observations of optical orphan afterglows having redshift measurements.
AT\,2020blt had only one detection, at 10\,GHz \citep{Ho2020d}.
PTF11agg had detailed observations with the VLA and CARMA, but no redshift measurement \citep{Cenko2013}.
The spectral luminosities of $10^{31}\,\erg\,\psec\,\phz$ at $\Delta t\approx10\,\days$ are typical of LGRBs \citep{Chandra2012}.

The 10\,GHz radio light curves of AT\,2021any and AT\,2021lfa, which are shown in Figure~\ref{fig:orphan-lcs}, exhibit evidence for relatively short-timescale variability.
Both show rapid fading from the first to second observation, and the light curve of AT\,2021any has an abrupt rebrightening at $\Delta t=20\,$d.
The early fading in \edit1{the light curve of} AT\,2021lfa represents a factor-of-two flux decrease in a few hours.
For AT\,2021any, the flux drops by a factor of four from 4.91\,d to 6.86\,d.
The rebrightening at $\Delta t=20\,\days$ represents a factor of 2--3 change in flux over seven days in the observer frame (two days in the rest frame); the fade represents a factor of 5--6 over the same timescale.

Interstellar scintillation can cause short-term variations at these frequencies, and we find that it likely makes a significant contribution to the light curve of AT\,2021any.
Scintillation results from small-scale inhomogeneities in the interstellar medium (ISM), which change the phase of an incoming wavefront.
As the Earth moves, the line of sight to a background source changes,
so the net effect is an observed change in flux.
The effect is greatest for sources observed at a frequency $\nu_\mathrm{obs}$ that is close to the transition frequency $\nu_0$, which separates strong scattering ($\nu_\mathrm{obs}<\nu_0$) from weak scattering ($\nu_\mathrm{obs}>\nu_0$).
Using the NE2001 model of the ISM \citep{Cordes2002},
we find that the positions of AT\,2021any and AT\,2021lfa have a transition frequency $\nu_0=15\,$GHz and $\nu_0=8\,$GHz, respectively.
So, the 10\,GHz light curve of AT\,2021any is very likely affected by scintillation.
The 10\,GHz light curve of AT\,2021lfa may not be: it is possible that the earliest emission represents a truly distinct emission component, such as a reverse shock \citep{Kulkarni1999,Sari1999,Harrison2001,Laskar2013,Perley2014_130427A,Laskar2016}.

The SED evolution of AT\,2021any is shown in Figure~\ref{fig:radio-sed}.
The data at all epochs are fainter with increasing frequency, suggesting that the emission is optically thin at all epochs (i.e., that the synchrotron self-absorption frequency $\nu_a < 8\,$GHz).
We fit a power law of the form $f_\nu = f_0 \nu^{\beta}$ to each epoch, and find significant changes with time. In particular, the spectral index appears to be steeper ($f_\nu \propto \nu^{-2}$) during the brightest parts of the light curve.
The spectral index is shallower ($\nu^{-0.5}$) between brightening episodes, which is a more typical spectral index for GRB afterglows at frequencies below the cooling frequency $\nu_a < \nu < \nu_c$ \citep{Granot2002_shape_spectral_breaks}.
If the brightening is due to scintillation, it may be that the changing spectral index is also due to the frequency-dependent effects of scintillation.

In the full SED evolution of AT\,2021lfa (Figure~\ref{fig:at2021lfa-sed}), which spans 7.83--103.58\,d,
we appear to observe the transition from optically thick to optically thin regimes.
During the epochs 7.83--15.72\,d, corresponding to the first peak and rise of the light curve,
the data appear to be self-absorbed. We measure a power-law index of $\beta=1.15\pm0.01$ at $\Delta t=7.83\,$d and an index of $\beta=1.86\pm0.02$ at $\Delta t=11.82\,$d.
By 21.71\,d, the index has become significantly more shallow: we measure $\beta=0.26\pm0.01$.
During the fading of the light curve, at 47.74\,d and 103.58\,d, we measure an optically thin spectral index of $\beta=-0.66\pm0.21$ and $\beta=-0.68\pm0.01$ (respectively), indicating that the self-absorption frequency has passed through the VLA observing bands, and that the cooling frequency lies above the VLA bands at all epochs of observation.

In conclusion, we do not find any clear differences between the multiwavelength properties of the orphan optical afterglows and the population of LGRBs discovered by high-energy satellites, \edit1{although we defer
detailed modeling of the multifrequency radio light curves to future work.}
So, at this stage we have no evidence of a population of optical afterglows that are distinct from LGRBs.
We address the implications in \S\ref{sec:discussion}.

\section{Discussion}
\label{sec:discussion}

In \S\ref{sec:comparison-grb},
we concluded that the optically discovered afterglows resemble the population of on-axis LGRBs in terms of their $\gamma$-ray to radio properties.
Although the events listed in Table~\ref{tab:summary} were not selected in a fully consistent way,
we can still draw valuable conclusions from the ratio of events with associated detected GRBs to those without.

\subsection{Comparison to LGRB Rate}

First we compare our detected afterglow rate to the LGRB rate. 
To do this, we adopt an approach similar to that of \citet{Cenko2013}: construct a mock catalog of LGRBs, adopt a light curve from the \emph{Swift}/BAT follow-up sample obtained with the P60 telescope \citep{Cenko2009}, then check how many of these events would have been discovered by us using ZTF by folding the light curve through the log of ZTF observations.

We created a log of all the observations in which we could have reliably discovered an afterglow.
To be conservative, we only used field-nights in which an event could have been recognized via intranight fading.
We used \texttt{ztfquery} and the ZTF observation log to select all field-nights in the years 2020 and 2021 with the following criteria:
\begin{enumerate}
    \item A typical limiting magnitude fainter than $20\,$mag.
    \item At least two $r$-band observations that night.
    \item An $r$-band observation the previous night in the same filter with a limiting magnitude fainter than $20\,$mag.
    \item A field with Galactic latitude $|b|>15\,$deg.
\end{enumerate}

\noindent Applying the criteria above resulted in an observation log of 19,190 field-nights and 35,171 unique observations.
We used \texttt{ztfquery} to estimate a typical limiting magnitude for each of the field-nights in each filter.

Next, we constructed a mock catalog of LGRBs. In the last few years, \emph{Swift}/BAT detected an average of 73\,LGRBs\,\pyr.
Accounting for the field of view and duty cycle in the same way as \citet{Cenko2013}, we estimate an all-sky rate of 511\,\pyr, slightly less than the rate of 630\,\pyr\ adopted by \citet{Cenko2013} (which included SGRBs).
In two years of ZTF, we therefore expect there to be 1022 LGRBs.
We assigned each of our 1022 mock GRBs a random burst time $t_0$, uniformly distributed between 1 January 2020 and 31 December 2021; a random R.A. and Decl. uniformly distributed across the sky; and a random GRB $R$-band light curve based on the sample of \citet{Cenko2009}.
We excluded four events --- GRB\,050607, GRB\,060110, GRB\,071003, and GRB\,071011 --- because they had Galactic latitudes $|b|<15^{\circ}$.
For the light curve, we obtained P60 $R_C$-band data from \citet{Cenko2009}, and in some cases added photometry from the literature \citep{Cenko2006_GRB,Soderberg2007,Perley2008,Covino2008,Littlejohns2012}.
If the data did not extend to below the ZTF detection threshold, we extrapolated the light curve using the best-fit power law from \citet{Cenko2013}.

\begin{figure*}[!htb]
    \centering
    \includegraphics[width=0.8\textwidth]{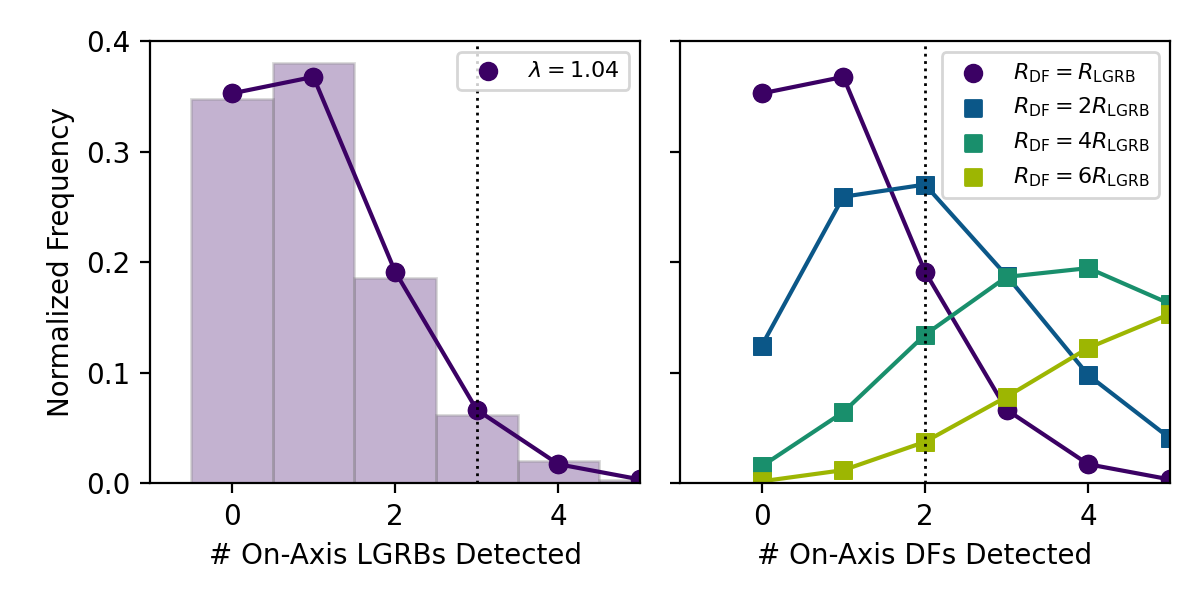}
    \caption{
    \emph{Left:} The simulated number of LGRB afterglows serendipitously detected by ZTF as intranight transients in two years of observations (2020 and 2021) based on the all-sky \emph{Swift}/BAT rate, the sample of \emph{Swift}/BAT afterglows from \citet{Cenko2009}, and 1000 Monte Carlo trials. The distribution is well described by a Poisson function with $\lambda=1.04$. The vertical dotted line shows the observed number, including one LGRB-associated and two orphan events.
    \emph{Right:} The simulated number of dirty fireballs serendipitously detected by ZTF for \edit1{four} different hypothetical relative rates. The vertical dotted line shows the number of observed orphan afterglows. If both are dirty fireballs, their rate does not exceed $6\times$ the LGRB rate (95\% confidence). 
    If neither are dirty fireballs, then the upper limit is $3\times$ the LGRB rate (95\% confidence).
    Note that the true number of ZTF-detected afterglows is higher; we consider a restricted sample for the rate estimate to ensure completeness.
    }
    \label{fig:sim}
\end{figure*}

We considered nine bursts ``dark'' to the P48\footnote{GRB\,050412, GRB\,050915A, GRB\,060510B, GRB\,060805A, GRB\,060923A, GRB\,061222A, GRB\,070521, GRB\,080320, and GRB\,050607.}.
For the remainder, we sampled the light curve using the ZTF observation log, to check if it would have been identified by us.
We defined a detection as being 0.5\,mag brighter than the limiting magnitude.
We checked if the set of detections met the following criteria:
\begin{enumerate}
    \item First detection brighter than 20\,mag.
    \item Pair of detections within the first night with separation $\Delta t$ where 0.02\,d $< \Delta t < 0.6\,$d.
\end{enumerate}

\noindent The above criteria were chosen to be a restrictive subset of our filter, so that we can be more confident in our completeness. Note that observations were grouped by field ID, so we ignore overlap between fields. Taking overlap into account would slightly increase the LGRB discovery rate.
Three afterglows passed these criteria in 2020 and 2021 (AT\,2020blt, AT\,2021any, and AT\,2021qbd), of which two (AT\,2020blt and AT\,2021any) were orphan.\footnote{An additional source, AT\,2021kym, passes the criteria if field overlap is taken into account.}

We ran the simulation 1000 times, and the resulting  number of expected detected LGRB afterglows is shown in the left panel of Figure~\ref{fig:sim}.
The vertical dotted line indicates the number of events we detected with ZTF under these criteria (which are more stringent than our actual criteria).
We estimate that the probability of detecting one LGRB afterglow is 60\%, while the probability of detecting three LGRB afterglows is only 7\%;
we can therefore be confident that our searches are reasonably complete.
Detecting three events appears unlikely, but we cannot formally rule it out, particularly given the uncertainties in the expected number of LGRBs, which we estimate to be $\sim10$\%.
Thus, from a rate estimate we do not have definitive evidence of a new class of relativistic explosions.

\subsection{The Rate of Dirty Fireballs}
\label{sec:discussion-df-rate}

We did not detect any confirmed new class of relativistic explosions using our experiment,
so can set the most robust upper limit to date on the dirty-fireball rate.
A longstanding puzzle in the LGRB field is the ``baryon loading problem'': that to efficiently produce $\gamma$-rays, the baryon loading content \edit1{must be} $M\lesssim10^{-4}\,\msol$ \citep{Piran2004}.
In the baryon-rich environment of a massive-star interior, it may be more natural for jets to become mass-loaded, in which case they could not accelerate material to $\Gamma_\mathrm{init}\gg100$ \citep{Huang2002}.
It has been argued that dirty fireballs would have an energy similar to that of clean fireballs (with the energy in lower-Lorentz factor material; \citealt{Huang2002}) and would be similarly collimated \citep{Rhoads2003}, in which case the optical afterglows would closely resemble each other \citep{Huang2002}.

Even in the extreme case that all three orphan afterglows (AT\,2020blt, AT\,2021any, AT\,2021lfa) are dirty fireballs, we can dismiss the possibility that dirty fireballs produce optical afterglows similar to those of LGRBs (as originally conceived) and are an order of magnitude more common than LGRBs.
This result is not surprising.
\citet{Cenko2015} and \citet{Ho2018} searched PTF data for extragalactic fast transients and recovered only one known GRB afterglow (iPTF14yb), already suggesting that dirty fireballs were not a significant population in the fast optical transient sky.
Searches for fast X-ray transients have also resulted in the conclusion that the ratio of dirty ($50 \lesssim \Gamma_0 \lesssim 200$) fireballs to LGRBs ($\Gamma_0 \gg 200$) can be no more than a factor of a few \citep{Grindlay1999,Dermer1999_dirty_fireballs,Greiner2000,Nakar2003}.

To make this limit quantitative, we use the result from our simulation in Figure~\ref{fig:sim}. The right panel of Figure~\ref{fig:sim} shows the expected number of detected on-axis dirty fireballs for \edit1{four} different hypothetical rates relative to the LGRB rate, assuming the same afterglow properties for both groups. The vertical dotted line displays the number of orphan afterglows detected (AT\,2020blt and AT\,2021any).
For each relative rate, we integrated the Poisson probability distribution function above $N=2$ to see when we would expect to see at least two events at 95\% confidence.
Assuming that both are dirty fireballs (which is unlikely), we rule out a scenario in which dirty fireballs produce similar on-axis optical afterglows to ordinary LGRBs and have a rate that is six times the LGRB rate (95\% confidence).
Assuming neither is a dirty fireball, which we consider more likely, the limit becomes three times the LGRB rate.

The lack of a large population of dirty fireballs has several possible explanations.
One possibility is that mass-loaded jets are less likely to escape the envelope.
Another is that the process of jet production and propagation somehow precludes the entrainment of matter.
A final possibility is that low-Lorentz-factor jets are successful but produce significantly different optical afterglows; for example, if they have a lower energy per solid angle, the optical afterglow would be fainter. \citet{Lei2013} investigated two different jet-launching mechanisms and found that baryon-rich jets tended to be less luminous.
A lower-Lorentz-factor outflow could also exhibit a longer plateau phase owing to the longer deceleration time \citep{Shen2012,Duffell2015}, resulting in suppressed early-afterglow emission.
As calculated by \citet{Ho2020d}, a dirty fireball with $\Gamma_\mathrm{init}=10$ would have a deceleration time of 1.2\,d.
A typical rest-frame $u$-band luminosity of LGRB afterglows at 1\,d in the rest frame is $10^{44}$\,\erg\,\psec\ \citep{Racusin2011}, about an order of magnitude fainter than the luminosity at which we are discovering afterglows (Figure~\ref{fig:lc-rest}). So, more sensitive searches may be required to test this scenario.

\subsection{The Optical Beaming Factor}

Finding optical afterglows without a GRB trigger is also of interest as a way to directly constrain the solid angle of the material producing optical afterglow emission,
sometimes referred to as the ``optical beaming factor,'' $f_{b,\mathrm{opt}}$.
As discussed by \citet{Nakar2003},
the collimation-corrected GRB energy is typically calculated using the optical beaming factor, with an implicit assumption that $f_{b,\mathrm{opt}}\approx f_{b,\mathrm{\gamma}}$ \citep{Frail2001,Panaitescu2001_clustering}.
However, until now there has been no direct test of this assumption.

Our work establishes that $f_{b,\mathrm{opt}}\approx f_{b,\mathrm{\gamma}}$.
By the same argument as in \S\ref{sec:discussion-df-rate},
we find $f_{b,\mathrm{opt}} < 6f_{b,\mathrm{\gamma}}$ at 95\% confidence, if our orphan events were viewed outside the $\gamma$-ray emitting region.
If the orphan events were simply LGRBs missed by satellites, then we have $f_{b,\mathrm{opt}} < 3f_{b,\mathrm{\gamma}}$.
This result is consistent with top-hat jet models (the ``spherical approximation''), because in these models, the beaming is expected to be similar for all of the relativistic material.

\citet{Nakar2003} performed a similar exercise and came to a \edit1{similar} conclusion, using X-ray afterglows.
They found that the X-ray beaming factor must be close to the $\gamma$-ray beaming factor, concluding that the bulk energies at $\Gamma=200$ (the $\gamma$-ray emitting material) and $\Gamma=10$ (the X-ray emitting material) are similar
and that the homogeneous-jet approximation is reasonable.
Our work shows that the energies in the $\gamma$-ray, X-ray, and optical-emitting material are all similar.

The result that the beaming factor of the afterglow-emitting material is similar to the beaming factor of the $\gamma$-ray-emitting material is consistent with jet structures predicted by simulations, and the emission predicted from analytical modeling.
In collapsar simulations, the jet develops radial and angular structure from its interaction with the dense stellar material \citep{Zhang2004_simulation,McKinney2006,Tchekhovskoy2008,Duffell2015,Gottlieb2021}. Most relevant for this work is the angular structure: a narrow ultrarelativistic core with a wide mildly relativistic sheath or cocoon, sometimes referred to as a ``two-component jet" (see \citealt{Granot2010} for a review).
The relative amount of energy in the wide and narrow components depends on a number of factors, but is likely at most comparable for successful jets \citep{RamirezRuiz2002,Nakar2017,deColle2018}.
So, the energy per solid angle from the wider component should be significantly lower than that of the narrow component, leading to fainter afterglow emission, as predicted by \citet{Nakar2017}.
We would therefore expect to be biased against afterglow emission from the wide-angle material in our current searches for cosmological ($z\gtrsim1$) transients.

\subsection{The Prevalence of Relativistic Jets in Collapsing Massive Stars}

By searching for cosmological relativistic explosions in the ZTF data, we have found no clear new phenomenon that is more common than LGRBs.
This result has implications for the fraction of CC~SNe that harbor central engines and successful, LGRB-energy relativistic outflows.
The observed rate of LGRBs has large uncertainties, as does the beaming fraction; from \edit1{LGRB} rates alone, the intrinsic LGRB rate could be anywhere from 0.01\% to 1\% of the CC~SN rate (see Table~10 of \citealt{Ho2020b}).
Radio searches have discovered one likely off-axis afterglow,
and have constrained the rate to be 40--240\,\pgpccub\,\pyr, or 0.06\%--0.1\% of the CC~SN rate \citep{Mooley2022}.
Our searches support the idea that energetic relativistic outflows are rare, i.e., that the rate is within a factor of a few of the LGRB rate.
However, we cannot set any constraints on the prevalence of weaker, lower-energy jets.

\section{Summary and Conclusions}
\label{sec:conclusions}

In this paper, \edit1{we presented the discovery of six cosmological fast optical transients} discovered by ZTF without a GRB trigger. \edit1{Our work doubles the number of optically discovered afterglows (for a total of 13 to date).}
Extragalactic fast transients powered by optically thin synchrotron emission can be efficiently discovered using high-cadence observations, by requiring rapid evolution (to rule out ordinary extragalactic transients like SNe) and red colors (to rule out the primary contaminant, stellar flares).
Using rapid-turnaround optical spectroscopy, X-ray, and radio observations, we measured the redshift of almost all of the events, and showed that they closely resemble on-axis LGRB afterglows.
\edit1{Of the ten afterglows discovered by ZTF to date,
six had an associated detected LGRB identified in a post facto search.
This result} rules out a scenario in which low-Lorentz-factor jets (``dirty'' or ``failed'' fireballs) have an energy per solid angle similar to that of clean fireballs and are \edit1{an order of magnitude} more common than classical GRBs, \edit1{which is} consistent with past searches at X-ray and optical wavelengths.
In addition, we set the first direct constraint on the optical beaming factor in LGRBs, finding that it must be comparable to the $\gamma$-ray beaming factor.

Our searches were originally motivated by the search for dirty fireballs.
The discovery of a population of optical afterglows with a rate greatly exceeding the LGRB rate would have lent strong support for their existence \citep{Cenko2013}.
However, it is now clear that it is not so simple.
It may be more efficient to search for the prompt X-ray emission expected to accompany a dirty fireball, such as the X-ray flashes found by HETE-2 \citep{Sakamoto2005}.
This will become possible in the next few years with facilities such as the Space Variable Object Monitor (SVOM; \citealt{Wei2016,Cordier2015}) and Einstein-Probe \citep{Yuan2018},
and perhaps in the next decade with facilities like the Gamow Explorer \citep{White2021}.
Millimeter-wavelength observations of the reverse shock could help infer the initial Lorentz factor (e.g., \citealt{Laskar2019}).

In addition, our current searches are not sensitive to fainter populations, such as highly off-axis afterglows or very dirty fireballs with a long deceleration time.
Finding off-axis events is essential for studying LGRB jet structure, and the lack of a significant population of luminous dirty fireballs simplifies the picture.
We will address search strategies for finding slower-evolving relativistic explosions in future studies.

Our work \edit1{has} resulted in the discovery of \edit1{three} orphan events of unknown origin, \edit1{two of which (AT\,2021any and AT\,2021lfa) are presented in this paper.}
Ruling out an associated detected LGRB is not straightforward,
and at present requires determining the pointing histories of different high-energy satellites; this work benefited from the coordination enabled by the IPN.
Based on the coverage and sensitivity of the different spacecraft,
the simplest explanation for the orphan events is that they were ordinary LGRBs for which the prompt emission was missed.
To determine whether this was truly the case, more sensitive high-energy observations would be required.
An all-sky facility with a fluence threshold equal to that of the BAT ($10^{-8}\,\erg\,\pcmsq$) could
rule out an $E_{\gamma,\mathrm{iso}}=10^{51}\,\erg$ GRB out to $z=3$, which includes all of the events in our sample.
A burst with $10^{50}\,\erg$ could be ruled out at $z\lesssim 1.5$.

\edit1{Although our results suggest that dirty fireballs are not significantly more common than classical GRBs,
a significantly larger sample of afterglows would be needed to determine whether a population of dirty fireballs exists at a rate less than or comparable to the GRB rate.}
With a much higher afterglow discovery rate, manual scanning and triggering will be impractical, as it already is for GRBs themselves. This would motivate an optical version of the \emph{Swift} facility: autonomous candidate identification and triggering for confirmation. Some of the events in our sample could have been intranight triggers, since there was a nondetection followed by a detection on the same night.
With a manageable false-positive rate, deep imaging facilities could confirm a candidate based on fade rate and colors. Spectroscopy to measure the redshift, and X-ray and radio follow-up observations to detect the afterglow at other wavelengths, could then be autonomously performed.
Given that the typical time from the close of the ZTF shutter to the release of alerts is 8\,min \citep{Masci2019}, with higher latencies primarily coming from crowded Galactic fields, it is not unreasonable to strive for afterglow identification, classification, and follow-up observations within 30\,min of core collapse.

\vspace{5mm}
\facilities{Swift, EVLA, VLA, Liverpool:2m, PO:1.2m, PO:1.5m, GTC, Gemini:South, Fermi, CAO:2.2m, Keck:I (LRIS)}

\software{{\tt CASA} \citep{McMullin2007},
          {\tt astropy} \citep{Astropy2013,Astropy2018},
          {\tt matplotlib} \citep{Hunter2007},
          {\tt scipy} \citep{Virtanen2020},
          {\tt ztfquery} (Rigault 2018),
          {\tt extinction}, {\tt penquins},
          {\tt pandas} \citep{reback2020pandas,mckinney-proc-scipy-2010}
}

\acknowledgements

We acknowledge with gratitude the contributions of the late Kevin Hurley in founding and maintaining the Interplanetary Network, which was essential for this work.

The authors would like to thank WeiKang Zheng at U.C. Berkeley for assistance with Keck observations.
A.Y.Q.H. would like to thank Ragnhild Lunnan for helpful comments on the manuscript; and Eliot Quataert, Dan Kasen, Andrew MacFadyen, and Paul Duffell for fruitful discussions about jet structure and dirty fireballs.
D.A.P.'s contribution was performed in part at the Aspen Center for Physics, which is supported by National Science Foundation (NSF) grant PHY-1607611.  This work was partially supported by a grant from the Simons Foundation.
D.F., A.T., and M.U. acknowledge support from RSF grant 21-12-00250.
D.A.K. and J.F.A.F acknowledges support from Spanish National Research Project RTI2018-098104-J-I00 (GRBPhot).
H.K. thanks the LSSTC Data Science Fellowship Program, which is funded by LSSTC, NSF Cybertraining Grant \#1829740, Brinson and Moore Foundations.
J.F.A.F. acknowledges support from the Spanish Ministerio de Ciencia, Innovación y Universidades through the grant PRE2018-086507.
A.V.F.'s group at U.C. Berkeley is grateful for assistance from the Christopher
R. Redlich Fund and many individual donors.

Based on observations obtained with the Samuel Oschin Telescope 48-inch and the 60-inch Telescope at the Palomar Observatory as part of the Zwicky Transient Facility project. ZTF is supported by the NSF under grants AST-1440341 and AST-2034437 and a collaboration including current partners Caltech, IPAC, the Weizmann Institute for Science, the Oskar Klein Center at Stockholm University, the University of Maryland, Deutsches Elektronen-Synchrotron and Humboldt University, the TANGO Consortium of Taiwan, the University of Wisconsin at Milwaukee, Trinity College Dublin, Lawrence Livermore National Laboratories, IN2P3, University of Warwick, Ruhr University Bochum, Northwestern University and former partners the University of Washington, Los Alamos National Laboratories, and Lawrence Berkeley National Laboratories. Operations are conducted by COO, IPAC, and UW. 

The Liverpool Telescope is operated on the island of La Palma by Liverpool John Moores University in the Spanish Observatorio del Roque de los Muchachos of the Instituto de Astrofisica de Canarias with financial support from the UK Science and Technology Facilities Council.

SED Machine is based upon work supported by the NSF under grant 1106171.

Based on observations obtained at the international Gemini Observatory, a program of NSF’s NOIRLab, which is managed by the Association of Universities for Research in Astronomy (AURA), Inc., under a cooperative agreement with the NSF on behalf of the Gemini Observatory partnership: the NSF (U.S.), National Research Council (Canada), Agencia Nacional de Investigaci\'{o}n y Desarrollo (Chile), Ministerio de Ciencia, Tecnolog\'{i}a e Innovaci\'{o}n (Argentina), Minist\'{e}rio da Ci\^{e}ncia, Tecnologia, Inova\c{c}\~{o}es e Comunica\c{c}\~{o}es (Brazil), and Korea Astronomy and Space Science Institute (Republic of Korea).

Partially based on observations made with the Gran Telescopio Canarias
(GTC), installed at the Spanish Observatorio del Roque de los Muchachos
of the Instituto de Astrof\'isica de Canarias, on the island of La
Palma.
Partially based on observations collected at the Centro Astron\'omico
Hispano en Andalucía (CAHA) at Calar Alto, operated jointly by Junta de
Andalucía and Consejo Superior de Investigaciones Científicas
(IAA-CSIC).
The National Radio Astronomy Observatory is a facility of the NSF operated under cooperative agreement by AURA, Inc.

Some of the data presented herein were obtained at the W. M. Keck Observatory, which is operated as a scientific partnership among the California Institute of Technology, the University of California, and the National Aeronautics and Space Administration. The Observatory was made possible by the generous financial support of the W. M. Keck Foundation.
The authors wish to recognize and acknowledge the very significant cultural role and reverence that the summit of Maunakea has always had within the indigenous Hawaiian community.  We are most fortunate to have the opportunity to conduct observations from this mountain.

\appendix

\section{Optical Photometry}
\label{sec:optical-photometry}

In Table~\ref{tab:opt-phot} we provide optical photometry for the afterglows in our sample.

\startlongtable 
\begin{deluxetable}{lrrrrr} 
\tablecaption{Optical photometry for afterglows, not corrected for Milky Way extinction. \label{tab:opt-phot}} 
\tablewidth{0pt} 
\tablehead{ \colhead{Name} & \colhead{Date} & \colhead{$\Delta t^\dag$} & \colhead{Inst.} & \colhead{Filt.} & \colhead{Mag} \\ \colhead{} & \colhead{(MJD)} & \colhead{(d)} & \colhead{} & \colhead{} & \colhead{(AB)}} 
\tabletypesize{\scriptsize} 
\startdata 
AT2021any & 59230.2916 & 0.0141 & P48$^\ddag$ & $r$ & $17.92 \pm 0.02$ \\
AT2021any & 59230.3307 & 0.0532 & P48 & $g$ & $19.35 \pm 0.05$ \\
AT2021any & 59230.3316 & 0.0541 & P48 & $g$ & $19.41 \pm 0.06$ \\
AT2021any & 59230.3563 & 0.0788 & P48 & $g$ & $19.67 \pm 0.06$ \\
AT2021any & 59230.3712 & 0.0937 & P48 & $r$ & $19.40 \pm 0.05$ \\
AT2021any & 59230.3717 & 0.0942 & P48 & $r$ & $19.41 \pm 0.05$ \\
AT2021any & 59230.4303 & 0.1528 & P48 & $r$ & $19.91 \pm 0.11$ \\
AT2021any & 59230.9772 & 0.6997 & GTC & $r$ & $21.74 \pm 0.08$ \\
AT2021any & 59232.0096 & 1.7321 & CAHA & $r$ & $22.75 \pm 0.13$ \\
AT2021any & 59233.0184 & 2.7409 & CAHA & $r$ & $22.92 \pm 0.12$ \\
AT2020kym & 58993.2863 & 0.0752 & P48 & $r$ & $17.33 \pm 0.01$ \\
AT2020kym & 58993.2886 & 0.0775 & P48 & $r$ & $17.38 \pm 0.02$ \\
AT2020kym & 58993.3041 & 0.0930 & P48 & $r$ & $17.63 \pm 0.02$ \\
AT2020kym & 58993.3065 & 0.0954 & P48 & $r$ & $17.66 \pm 0.02$ \\
AT2020kym & 58994.2141 & 1.0029 & P48 & $g$ & $21.60 \pm 0.21$ \\
AT2020kym & 58994.2993 & 1.0882 & P48 & $r$ & $21.33 \pm 0.21$ \\
AT2021cwd & 59257.3697 & 0.2600 & P48 & $g$ & $19.57 \pm 0.09$ \\
AT2021cwd & 59257.4158 & 0.3061 & P48 & $r$ & $19.51 \pm 0.07$ \\
AT2021cwd & 59257.9850 & 0.8753 & LT & $g$ & $21.65 \pm 0.17$ \\
AT2021cwd & 59257.9860 & 0.8763 & LT & $r$ & $20.88 \pm 0.13$ \\
AT2021cwd & 59257.9871 & 0.8774 & LT & $i$ & $20.80 \pm 0.13$ \\
AT2021cwd & 59257.9895 & 0.8798 & LT & $z$ & $20.53 \pm 0.20$ \\
AT2021cwd & 59258.9459 & 1.8362 & LT & $g$ & $23.32 \pm 0.32$ \\
AT2021cwd & 59258.9509 & 1.8412 & LT & $r$ & $22.42 \pm 0.21$ \\
AT2021cwd & 59258.9559 & 1.8462 & LT & $i$ & $22.57 \pm 0.27$ \\
AT2021lfa & 59338.2324 & 0.9393 & P48 & $r$ & $18.60 \pm 0.08$ \\
AT2021lfa & 59338.3126 & 1.0195 & P48 & $g$ & $18.80 \pm 0.11$ \\
AT2021lfa & 59338.8893 & 1.5962 & LT & $g$ & $20.12 \pm 0.04$ \\
AT2021lfa & 59338.8920 & 1.5988 & LT & $r$ & $19.75 \pm 0.03$ \\
AT2021lfa & 59338.8946 & 1.6015 & LT & $i$ & $19.46 \pm 0.04$ \\
AT2021lfa & 59339.0342 & 1.7411 & LT & $u$ & $21.73 \pm 0.31$ \\
AT2021lfa & 59339.0376 & 1.7445 & LT & $g$ & $20.52 \pm 0.04$ \\
AT2021lfa & 59339.0403 & 1.7472 & LT & $r$ & $20.12 \pm 0.04$ \\
AT2021lfa & 59339.0429 & 1.7498 & LT & $i$ & $19.82 \pm 0.04$ \\
AT2021lfa & 59339.0456 & 1.7525 & LT & $z$ & $19.63 \pm 0.07$ \\
AT2021lfa & 59339.1898 & 1.8967 & SEDM & $g$ & $20.88 \pm 0.08$ \\
AT2021lfa & 59339.1925 & 1.8994 & SEDM & $r$ & $20.41 \pm 0.07$ \\
AT2021lfa & 59339.1952 & 1.9021 & SEDM & $i$ & $20.05 \pm 0.08$ \\
AT2021lfa & 59339.8817 & 2.5886 & LT & $g$ & $21.70 \pm 0.06$ \\
AT2021lfa & 59339.8868 & 2.5936 & LT & $r$ & $21.19 \pm 0.05$ \\
AT2021lfa & 59339.8917 & 2.5986 & LT & $i$ & $21.09 \pm 0.07$ \\
AT2021lfa & 59340.8923 & 3.5992 & LT & $g$ & $22.36 \pm 0.08$ \\
AT2021lfa & 59340.8973 & 3.6042 & LT & $r$ & $22.10 \pm 0.10$ \\
AT2021lfa & 59340.9023 & 3.6092 & LT & $i$ & $21.77 \pm 0.10$ \\
AT2021lfa & 59341.8829 & 4.5897 & LT & $g$ & $23.10 \pm 0.14$ \\
AT2021lfa & 59341.8878 & 4.5947 & LT & $r$ & $22.83 \pm 0.14$ \\
AT2021lfa & 59341.8928 & 4.5997 & LT & $i$ & $22.34 \pm 0.15$ \\
AT2021lfa & 59343.9336 & 6.6405 & LT & $g$ & $24.04 \pm 0.28$ \\
AT2021lfa & 59343.9419 & 6.6488 & LT & $r$ & $22.88 \pm 0.14$ \\
AT2021lfa & 59343.9501 & 6.6570 & LT & $i$ & $22.89 \pm 0.23$ \\
AT2021qbd & 59376.2325 & 0.4053 & P48 & $g$ & $18.49 \pm 0.02$ \\
AT2021qbd & 59376.2752 & 0.4480 & P48 & $r$ & $18.23 \pm 0.02$ \\
AT2021qbd & 59376.3206 & 0.4935 & P48 & $r$ & $18.37 \pm 0.02$ \\
AT2021qbd & 59376.3601 & 0.5329 & P48 & $g$ & $18.77 \pm 0.03$ \\
AT2021qbd & 59377.2694 & 1.4422 & P48 & $r$ & $20.30 \pm 0.10$ \\
AT2021qbd & 59377.3326 & 1.5055 & P48 & $g$ & $20.75 \pm 0.14$ \\
AT2021qbd & 59377.3336 & 1.5064 & P48 & $g$ & $20.87 \pm 0.12$ \\
AT2021qbd & 59377.3613 & 1.5341 & P48 & $g$ & $20.80 \pm 0.14$ \\
AT2021qbd & 59377.3959 & 1.5687 & P48 & $r$ & $20.24 \pm 0.10$ \\
AT2021qbd & 59378.2548 & 2.4277 & P48 & $r$ & $21.69 \pm 0.33$ \\
AT2021qbd & 59378.3232 & 2.4961 & P48 & $r$ & $21.14 \pm 0.20$ \\
\enddata 
\tablenotetext{\dag}{Time given relative to $t_0$ as estimated in the text.}
\tablenotetext{\ddag}{P48 values were measured using forced photometry \citep{Yao2019}.}
\end{deluxetable} 

\onecolumngrid

\newpage

\section{X-ray Observations}

In Table~\ref{tab:xray-phot} we provide a log of our X-ray observations of the orphan afterglows AT\,2021any and AT\,2021lfa.

\begin{deluxetable*}{lrrrrr}[!h]
\tablecaption{0.3--10\,keV X-ray observations for afterglows from \emph{Swift}/XRT. \label{tab:xray-phot}} 
\tablewidth{0pt} 
\tablehead{ \colhead{Name} & \colhead{Date} & \colhead{$\Delta t^\dag$} & \colhead{Exp.} & \colhead{Count Rate} & \colhead{Flux$^{\ddag\ddag}$} \\[-0.3cm] \colhead{} & \colhead{(MJD)} & \colhead{(d)} & \colhead{(ks)} & \colhead{($10^{-3}\,\psec$)} & \colhead{($10^{-13}\,\erg\,\psec\,\pcmsq$)}} 
\tabletypesize{\scriptsize} 
\startdata 
AT\,2021any & 59231.10 & 0.83 & 3.7 & $7.20 \pm 2.00^{\ddag}$ & $3.30 \pm 0.90$ \\
AT\,2021any & 59234.08 & 3.80 & 2.6 & $<4.70^{\dag\dag}$ & $<2.10$ \\
AT\,2021any & 59239.58 & 9.31 & 2.9 & $<4.30$ & $<1.90$ \\
AT\,2021lfa & 59339.23 & 1.94 & 5.0 & $9.30 \pm 1.70$ & $3.50 \pm 0.60$ \\
AT\,2021lfa & 59341.69 & 4.39 & 5.1 & $<4.40$ & $<1.60$ \\
\enddata 
\tablenotetext{\dag}{Time given relative to $t_0$ as estimated in the text.}
\tablenotetext{\ddag\ddag}{The conversion from X-ray count rate to unabsorbed flux uses a hydrogen column density listed in the text, and a photon index of $\Gamma=2$ for all sources.}
\tablenotetext{\ddag}{Uncertainties are 1$\sigma$.}
\tablenotetext{\dag\dag}{Upper limits are 3$\sigma$.}
\end{deluxetable*} 

\onecolumngrid

\section{Radio Observations}

In Table~\ref{tab:radio-phot} we provide a log of our VLA radio observations of the orphan afterglows AT\,2021any and AT\,2021lfa.

\startlongtable\begin{deluxetable*}{lrrrrrrr} 
\tablecaption{Log of our VLA radio observations of two afterglows with no associated detected GRBs.$^{\ddag\ddag}$\label{tab:radio-phot}} 
\tablewidth{0pt} 
\tablehead{ \colhead{Name} & \colhead{Date} & \colhead{$\Delta t^\dag$} & \colhead{Band} & \colhead{$\nu$} & \colhead{$\Delta \nu$} & \colhead{$F_\nu$} & \colhead{\edit1{Configuration}} \\[-0.3cm] \colhead{} & \colhead{(MJD)} & \colhead{(d)} & \colhead{} & \colhead{(GHz)} & \colhead{(GHz)} & \colhead{($\mu$Jy)} & \colhead{}} 
\tabletypesize{\scriptsize} 
\startdata 
AT2021any & 59235.19 & 4.91 & X & 9.981 & 3.58 & $91\pm5^\ddag$ & A \\
AT2021any & 59235.19 & 4.91 & X & 10.98 & 1.79 & $63\pm6$ & A \\
AT2021any & 59235.19 & 4.91 & X & 9.0 & 1.79 & $116\pm8$ & A \\
AT2021any & 59235.19 & 4.91 & X & 11.5 & 0.896 & $66\pm9$ & A \\
AT2021any & 59235.19 & 4.91 & X & 10.49 & 0.896 & $62\pm7$ & A \\
AT2021any & 59235.19 & 4.91 & X & 9.51 & 0.896 & $99\pm8$ & A \\
AT2021any & 59235.19 & 4.91 & X & 8.49 & 0.896 & $133\pm9$ & A \\
AT2021any & 59237.13 & 6.86 & X & 9.981 & 3.58 & $25\pm4$ & A \\
AT2021any & 59237.13 & 6.86 & X & 10.98 & 1.79 & $33\pm6$ & A \\
AT2021any & 59237.13 & 6.86 & X & 9.0 & 1.79 & $<15^{\dag\dag}$ & A \\
AT2021any & 59237.13 & 6.86 & X & 11.5 & 0.896 & $37\pm9$ & A \\
AT2021any & 59237.13 & 6.86 & X & 10.49 & 0.896 & $35\pm8$ & A \\
AT2021any & 59237.13 & 6.86 & X & 9.51 & 0.896 & $29\pm8$ & A \\
AT2021any & 59237.13 & 6.86 & Ku & 15.1 & 5.38 & $21\pm4$ & A \\
AT2021any & 59237.13 & 6.86 & Ku & 13.58 & 2.69 & $20\pm6$ & A \\
AT2021any & 59237.13 & 6.86 & Ku & 16.62 & 2.69 & $24\pm6$ & A \\
AT2021any & 59237.13 & 6.86 & Ku & 12.81 & 1.34 & $25\pm9$ & A \\
AT2021any & 59237.13 & 6.86 & Ku & 14.31 & 1.34 & $20\pm7$ & A \\
AT2021any & 59237.13 & 6.86 & Ku & 15.85 & 1.34 & $32\pm8$ & A \\
AT2021any & 59240.17 & 9.90 & S & 3.0 & 1.1 & $67\pm14$ & A \\
AT2021any & 59240.17 & 9.90 & S & 3.3 & 0.64 & $38\pm11$ & A \\
AT2021any & 59240.17 & 9.90 & S & 2.7 & 0.64 & $<47$ & A \\
AT2021any & 59240.17 & 9.90 & C & 5.1 & 1.8 & $<22$ & A \\
AT2021any & 59240.17 & 9.90 & C & 7.0 & 2.0 & $34\pm7$ & A \\
AT2021any & 59240.17 & 9.90 & C & 6.1 & 3.8 & $30\pm5$ & A \\
AT2021any & 59240.17 & 9.90 & X & 10.0 & 4.0 & $<21$ & A \\
AT2021any & 59240.17 & 9.90 & X & 9.0 & 2.0 & $<27$ & A \\
AT2021any & 59240.17 & 9.90 & Ku & 15.3 & 5.0 & $31\pm6$ & A \\
AT2021any & 59240.17 & 9.90 & Ku & 15.1 & 2.0 & $46\pm9$ & A \\
AT2021any & 59244.31 & 14.04 & X & 9.5 & 3.1 & $24\pm5$ & A \\
AT2021any & 59244.31 & 14.04 & X & 9.0 & 2.0 & $32\pm4$ & A \\
AT2021any & 59244.31 & 14.04 & X & 10.0 & 2.0 & $16\pm4$ & A \\
AT2021any & 59244.31 & 14.04 & X & 8.5 & 1.0 & $34\pm6$ & A \\
AT2021any & 59244.31 & 14.04 & X & 9.5 & 1.0 & $32\pm6$ & A \\
AT2021any & 59244.31 & 14.04 & X & 10.5 & 1.0 & $20\pm6$ & A \\
AT2021any & 59244.31 & 14.04 & X & 11.5 & 1.0 & $<24$ & A \\
AT2021any & 59251.08 & 20.81 & X & 9.7 & 3.6 & $78\pm6$ & A \\
AT2021any & 59251.08 & 20.81 & X & 9.0 & 2.0 & $89\pm6$ & A \\
AT2021any & 59251.08 & 20.81 & X & 10.8 & 1.6 & $67\pm9$ & A \\
AT2021any & 59251.08 & 20.81 & X & 8.5 & 1.0 & $98\pm8$ & A \\
AT2021any & 59251.08 & 20.81 & X & 9.5 & 1.0 & $78\pm7$ & A \\
AT2021any & 59251.08 & 20.81 & X & 10.5 & 1.0 & $72\pm11$ & A \\
AT2021any & 59251.08 & 20.81 & X & 11.5 & 1.0 & $43\pm15$ & A \\
AT2021any & 59258.25 & 27.97 & X & 9.7 & 3.6 & $13\pm4$ & A \\
AT2021any & 59273.11 & 42.84 & X & 9.7 & 3.6 & $18\pm3$ & A \\
AT2021any & 59306.06 & 75.78 & X & 9.0 & 2.0 & $<12$ & D \\
AT2021any & 59306.06 & 75.78 & X & 10.8 & 1.6 & $<17$ & D \\
AT2021lfa & 59340.01 & 2.72 & X & 9.0 & 2.0 & $46\pm7$ & D \\
AT2021lfa & 59340.01 & 2.72 & X & 10.8 & 1.6 & $<39$ & D \\
AT2021lfa & 59345.08 & 7.79 & X & 9.0 & 2.0 & $254\pm14$ & D \\
AT2021lfa & 59345.08 & 7.79 & X & 10.8 & 1.6 & $279\pm17$ & D \\
AT2021lfa & 59345.12 & 7.83 & Ku & 13.1 & 0.9 & $238\pm16$ & D \\
AT2021lfa & 59345.12 & 7.83 & Ku & 14.0 & 0.9 & $280\pm17$ & D \\
AT2021lfa & 59345.12 & 7.83 & Ku & 14.9 & 0.9 & $288\pm16$ & D \\
AT2021lfa & 59345.12 & 7.83 & Ku & 15.8 & 0.9 & $287\pm17$ & D \\
AT2021lfa & 59345.12 & 7.83 & Ku & 16.7 & 0.9 & $322\pm18$ & D \\
AT2021lfa & 59345.12 & 7.83 & Ku & 17.7 & 0.9 & $276\pm18$ & D \\
AT2021lfa & 59345.12 & 7.83 & X & 9.0 & 2.0 & $147\pm10$ & D \\
AT2021lfa & 59345.12 & 7.83 & X & 10.8 & 1.6 & $168\pm10$ & D \\
AT2021lfa & 59349.11 & 11.82 & Ku & 13.1 & 0.9 & $206\pm15$ & D \\
AT2021lfa & 59349.11 & 11.82 & Ku & 14.0 & 0.9 & $206\pm15$ & D \\
AT2021lfa & 59349.11 & 11.82 & Ku & 14.9 & 0.9 & $223\pm16$ & D \\
AT2021lfa & 59349.11 & 11.82 & Ku & 15.8 & 0.9 & $226\pm16$ & D \\
AT2021lfa & 59349.11 & 11.82 & Ku & 16.7 & 0.9 & $264\pm18$ & D \\
AT2021lfa & 59349.11 & 11.82 & Ku & 17.7 & 0.9 & $257\pm19$ & D \\
AT2021lfa & 59349.11 & 11.82 & X & 9.0 & 2.0 & $77\pm7$ & D \\
AT2021lfa & 59349.11 & 11.82 & X & 10.8 & 1.6 & $107\pm8$ & D \\
AT2021lfa & 59353.01 & 15.72 & Ku & 13.1 & 0.9 & $247\pm16$ & D \\
AT2021lfa & 59353.01 & 15.72 & Ku & 14.0 & 0.9 & $237\pm15$ & D \\
AT2021lfa & 59353.01 & 15.72 & Ku & 14.9 & 0.9 & $277\pm17$ & D \\
AT2021lfa & 59353.01 & 15.72 & Ku & 15.8 & 0.9 & $263\pm17$ & D \\
AT2021lfa & 59353.01 & 15.72 & Ku & 16.7 & 0.9 & $287\pm18$ & D \\
AT2021lfa & 59353.01 & 15.72 & Ku & 17.7 & 0.9 & $271\pm18$ & D \\
AT2021lfa & 59353.01 & 15.72 & X & 9.0 & 2.0 & $155\pm9$ & D \\
AT2021lfa & 59353.01 & 15.72 & X & 10.8 & 1.6 & $127\pm8$ & D \\
AT2021lfa & 59359.00 & 21.71 & Ku & 13.1 & 0.9 & $149\pm12$ & D \\
AT2021lfa & 59359.00 & 21.71 & Ku & 14.0 & 0.9 & $196\pm13$ & D \\
AT2021lfa & 59359.00 & 21.71 & Ku & 14.9 & 0.9 & $224\pm13$ & D \\
AT2021lfa & 59359.00 & 21.71 & Ku & 15.8 & 0.9 & $208\pm13$ & D \\
AT2021lfa & 59359.00 & 21.71 & Ku & 16.7 & 0.9 & $239\pm14$ & D \\
AT2021lfa & 59359.00 & 21.71 & Ku & 17.7 & 0.9 & $234\pm16$ & D \\
AT2021lfa & 59359.00 & 21.71 & X & 9.0 & 2.0 & $194\pm11$ & D \\
AT2021lfa & 59359.00 & 21.71 & X & 10.8 & 1.6 & $209\pm12$ & D \\
AT2021lfa & 59385.03 & 47.74 & X & 9.0 & 2.0 & $194\pm10$ & DtoC \\
AT2021lfa & 59385.03 & 47.74 & X & 10.8 & 1.6 & $172\pm10$ & DtoC \\
AT2021lfa & 59440.03 & 102.73 & L & 1.43 & 0.4 & $<585$ & C \\
AT2021lfa & 59440.03 & 102.73 & L & 1.81 & 0.4 & $<276$ & C \\
AT2021lfa & 59440.03 & 102.73 & C & 4.5 & 1.0 & $<42$ & C \\
AT2021lfa & 59440.03 & 102.73 & C & 5.5 & 1.0 & $84\pm12$ & C \\
AT2021lfa & 59440.03 & 102.73 & C & 6.5 & 1.0 & $60\pm11$ & C \\
AT2021lfa & 59440.03 & 102.73 & C & 7.5 & 1.0 & $47\pm10$ & C \\
AT2021lfa & 59440.03 & 102.73 & C & 5.0 & 2.0 & $54\pm13$ & C \\
AT2021lfa & 59440.03 & 102.73 & C & 7.0 & 2.0 & $62\pm9$ & C \\
AT2021lfa & 59440.87 & 103.58 & Ku & 13.1 & 2.0 & $64\pm7$ & C \\
AT2021lfa & 59440.87 & 103.58 & Ku & 15.1 & 2.0 & $60\pm6$ & C \\
AT2021lfa & 59440.87 & 103.58 & Ku & 17.1 & 2.0 & $45\pm8$ & C \\
AT2021lfa & 59440.87 & 103.58 & X & 8.5 & 1.0 & $74\pm8$ & C \\
AT2021lfa & 59440.87 & 103.58 & X & 9.5 & 1.0 & $60\pm8$ & C \\
AT2021lfa & 59440.87 & 103.58 & X & 10.5 & 1.0 & $41\pm8$ & C \\
AT2021lfa & 59440.87 & 103.58 & X & 11.5 & 1.0 & $54\pm11$ & C \\
AT2021lfa & 59440.87 & 103.58 & X & 9.0 & 2.0 & $67\pm6$ & C \\
AT2021lfa & 59440.87 & 103.58 & X & 11.0 & 2.0 & $45\pm6$ & C \\
AT2021lfa & 59440.87 & 103.58 & C & 4.5 & 1.0 & $140\pm13$ & C \\
AT2021lfa & 59440.87 & 103.58 & C & 5.5 & 1.0 & $95\pm11$ & C \\
AT2021lfa & 59440.87 & 103.58 & C & 6.5 & 1.0 & $49\pm8$ & C \\
AT2021lfa & 59440.87 & 103.58 & C & 7.5 & 1.0 & $67\pm8$ & C \\
AT2021lfa & 59440.87 & 103.58 & C & 5.0 & 2.0 & $118\pm11$ & C \\
AT2021lfa & 59440.87 & 103.58 & C & 7.0 & 2.0 & $59\pm7$ & C \\
AT2021lfa & 59509.61 & 172.32 & Ku & 12.5 & 1.0 & $31\pm11$ & B \\
AT2021lfa & 59509.61 & 172.32 & Ku & 13.5 & 1.0 & $49\pm7$ & B \\
AT2021lfa & 59509.61 & 172.32 & Ku & 14.5 & 1.0 & $29\pm8$ & B \\
AT2021lfa & 59509.61 & 172.32 & Ku & 15.5 & 1.0 & $26\pm8$ & B \\
AT2021lfa & 59509.61 & 172.32 & Ku & 16.5 & 1.0 & $22\pm8$ & B \\
AT2021lfa & 59509.61 & 172.32 & Ku & 17.5 & 1.0 & $46\pm10$ & B \\
AT2021lfa & 59509.61 & 172.32 & Ku & 13.1 & 2.0 & $42\pm7$ & B \\
AT2021lfa & 59509.61 & 172.32 & Ku & 15.1 & 2.0 & $27\pm5$ & B \\
AT2021lfa & 59509.61 & 172.32 & Ku & 17.1 & 2.0 & $32\pm6$ & B \\
AT2021lfa & 59509.61 & 172.32 & X & 9.0 & 2.0 & $38\pm5$ & B \\
AT2021lfa & 59509.61 & 172.32 & X & 11.0 & 2.0 & $28\pm5$ & B \\
AT2021lfa & 59509.61 & 172.32 & X & 10.0 & 4.1 & $33\pm4$ & B \\
AT2021lfa & 59509.61 & 172.32 & C & 5.0 & 2.0 & $50\pm8$ & B \\
AT2021lfa & 59509.61 & 172.32 & C & 7.0 & 2.0 & $28\pm7$ & B \\
AT2021lfa & 59517.54 & 180.24 & C & 5.0 & 2.0 & $24\pm8$ & B \\
AT2021lfa & 59517.54 & 180.24 & C & 7.0 & 2.0 & $37\pm7$ & B \\
AT2021lfa & 59582.39 & 245.10 & Ku & 15.1 & 6.1 & $17\pm3$ & B \\
AT2021lfa & 59582.39 & 245.10 & Ku & 13.6 & 3.1 & $21\pm5$ & B \\
AT2021lfa & 59582.39 & 245.10 & Ku & 16.6 & 3.1 & $13\pm5$ & B \\
AT2021lfa & 59582.39 & 245.10 & Ku & 12.8 & 1.5 & $27\pm8$ & B \\
AT2021lfa & 59582.39 & 245.10 & Ku & 14.3 & 1.5 & $14\pm6$ & B \\
AT2021lfa & 59582.39 & 245.10 & Ku & 15.8 & 1.5 & $8\pm7$ & B \\
AT2021lfa & 59582.39 & 245.10 & Ku & 17.4 & 1.5 & $17\pm8$ & B \\
AT2021lfa & 59582.39 & 245.10 & X & 10.0 & 4.1 & $19\pm4$ & B \\
AT2021lfa & 59582.39 & 245.10 & X & 9.0 & 2.0 & $22\pm5$ & B \\
AT2021lfa & 59582.39 & 245.10 & X & 11.0 & 2.0 & $16\pm6$ & B \\
AT2021lfa & 59582.39 & 245.10 & C & 6.1 & 3.8 & $30\pm5$ & B \\
AT2021lfa & 59582.39 & 245.10 & C & 7.0 & 2.0 & $23\pm6$ & B \\
AT2021lfa & 59582.39 & 245.10 & C & 5.1 & 1.8 & $32\pm7$ & B \\
\enddata 
\tablenotetext{\dag}{Time given relative to $t_0$ as estimated in the text.}
\tablenotetext{\ddag}{Uncertainties in radio measurements are given as the quadrature sum of the image RMS and a 5\% uncertainty in the flux density owing to flux calibration. }
\tablenotetext{\dag\dag}{Radio upper limits are $3\times$ the image root-mean square (RMS).}
\tablenotetext{\ddag\ddag}{Some measurements are not independent, i.e., the same bandpass is broken into windows and also quoted as the full bandwidth together.}
\end{deluxetable*}

\bibliography{refs}{}
\bibliographystyle{aasjournal}

\end{document}